\documentclass[12pt]{article}
\usepackage{fullpage}
\usepackage{cite}
\usepackage{amsmath}
\usepackage{amstext,amsfonts,amsbsy,eucal,amssymb}
\usepackage{amsmath,pifont}
\usepackage{color}
\usepackage{xypic}
\xyoption{all}
\xyoption{knot}
\usepackage{epsfig,psfrag}

\usepackage{times}
\newfont{\cyr}{wncyr10}
\numberwithin{equation}{section}
\renewcommand{\thefootnote}{\fnsymbol{footnote}}

\def\openone{\leavevmode\hbox{\small1\kern-3.8pt\normalsize1}}%
\DeclareMathOperator{\sh}{sh}

\DeclareMathOperator{\Tr}{Tr}

\renewcommand{\Im}{\text{Im}}

\newcommand{\Li}{\mathrm{Li}_2}
\begin{document}

\baselineskip 21pt
\parskip 7pt

\hfill  December 23, 2000

\hfill (Revised Edition)
\vskip -1cm

\vspace{24pt}

\begin{center}

  {\Large\textbf{
      Hyperbolic Structure Arising from a Knot Invariant
      }
    }

  \vspace{24pt}

  {\large Kazuhiro \textsc{Hikami}}
  \footnote[2]{E-mail:
    \texttt{hikami@phys.s.u-tokyo.ac.jp}
    }

  \textsl{Department of Physics, Graduate School of Science,\\
    University of Tokyo,\\
    Hongo 7--3--1, Bunkyo, Tokyo 113--0033, Japan.
    }

(Received: \hspace{40mm})

\end{center}


\begin{center}
  \underline{ABSTRACT}
\end{center}

We study the knot invariant based on the quantum dilogarithm function.
This invariant can  be regarded as a non-compact analogue of Kashaev's
invariant, or the
colored Jones invariant, and is defined  by an integral form.
The 3-dimensional picture of our invariant
originates from  the pentagon identity of
the quantum dilogarithm function,
and
we show  that  the hyperbolicity consistency conditions in gluing
polyhedra
arise naturally
in the classical limit
as the saddle point equation of our invariant.

\vfill
\noindent
\textsf{Key Words:}


\newpage

\renewcommand{\thefootnote}{\arabic{footnote}}

\section{Introduction}

Since
the discovery of  the Jones polynomial~\cite{Jones85},
many knot invariants are proposed.
In  construction of these  quantum invariants, the quantum group
plays a crucial role,
and a representation of the
braid generator is derived  from  the
universal $R$-matrix ~\cite{Turae88}.
Contrary to
that the Alexander polynomial was known to be related with the
homology of the universal abelian covering,
the quantum invariants  still lack
the geometrical interpretation.

In Ref.~\citen{Kasha95}, Kashaev introduced 
the knot invariant   by
use of the finite dimensional representation of  the quantum
dilogarithm function.
He further  conjectured~\cite{Kasha96b} that the asymptotic behavior of
this invariant for a hyperbolic   knot $K$ gives  the hyperbolic volume of
the knot complement $S^3 \setminus K$.
As it is well known that the hyperbolic volume of the ideal
tetrahedron is given by the Lobachevsky function~\cite{WPThurs80Lecture,JMiln82a} which
is closely related with the dilogarithm function,
his  conjecture
may sound  natural.
Later in Ref.~\citen{MuraMura99a}
Kashaev's  knot invariant was shown to be equivalent with the colored Jones
polynomial at  a specific value, and his
conjecture is rewritten as the ``volume conjecture'';
\begin{equation}
  \label{volume_conjecture}
  \| K \|
  =
  \frac{1}{v_3}
  \lim_{N\to \infty}
  \frac{2 \, \pi}{N} \log | J_N(K) | ,
\end{equation}
where  $\| K \|$ is the Gromov norm of $S^3 \setminus K$, and
$v_3$ is the hyperbolic volume of the regular ideal tetrahedron.
The knot invariant  $J_N(K)$ is defined from the colored Jones
polynomial
$V_N(K;t)$
($N$-dimensional representation of $s\ell_2$)
by
\begin{equation*}
  J_N(K) 
  =
  V_N(K ; \mathrm{e}^{\frac{2 \, \pi \, \mathrm{i}}{N}}) .
\end{equation*}
Thus to clarify  a geometrical property of the quantum knot invariants
such as the Jones polynomial,
it is very fascinating to  reveal the 3-dimensional picture of this
invariant.
Recently some geometrical aspects for the
conjecture~\eqref{volume_conjecture}
have been proposed  in
Refs.~\citen{YYokot00b}
(see also Ref.~\cite{HMuraka00c}) based on the 3-dimensional picture
of Ref.~\citen{DThurs99a}.

In this paper we define the knot invariant
as a ``non-compact'' analogue of Kashaev's invariant, or the colored
Jones invariant.
This is based on an infinite dimensional representation of the quantum
dilogarithm function, and
both the $R$-matrix and  the invariant  are  defined in an integral form.
In our construction a    parameter $\gamma$ which corresponds to
$\pi/N$ in eq.~\eqref{volume_conjecture} is regarded as
the Planck constant $\hbar/2$, and a limit in
eq.~\eqref{volume_conjecture} is realized by the classical limit
$\gamma \to 0$.
We shall  demonstrate how the hyperbolic structure
appears in the classical limit of the non-compact Jones invariant.

This paper is organized as follows.
In \S~\ref{sec:dilog} we review the properties of the classical and
quantum dilogarithm functions.
A key is that both functions satisfy the so-called pentagon identity.
Using these properties
we   construct a solution of the Yang--Baxter equation in terms
of the quantum dilogarithm function.
With this  $R$-operator, we introduce  the knot invariant in
\S~\ref{sec:invariant}.
This invariant is given  in the integral form from the beginning.
We recall that the integral form of the quantum dilogarithm function
was used in Ref.~\citen{Kasha96b} to elucidate the asymptotic
behavior of the colored Jones polynomial.
In \S~\ref{sec:asymptotic} we show that the hyperbolic structure
naturally appears in the classical limit of  our knot invariant.
We find that
in  $\gamma\to 0$ limit
the oriented ideal tetrahedron with transverse oriented faces is
associated to  the matrix elements of the quantum dilogarithm
function.
Correspondingly the $R$-operator is identified with the oriented
octahedron, whose vertices belong to the link $L$.
This explains how the octahedron was introduced for each braiding  in
Ref.~\citen{DThurs99a}.
We can apply the saddle point method to evaluate the asymptotic
behavior of the classical limit of the knot invariant, and
we further demonstrate that the   saddle point equation for  integrals
of the knot invariant exactly coincides with the hyperbolicity
consistency condition in gluing faces.
Combining the fact that the imaginary part of the classical
dilogarithm function gives the hyperbolic volume of the ideal
tetrahedron at the critical point,
we can  conclude that the  invariant is related with the hyperbolic
volume of the knot complement
at the critical point.
In \S~\ref{sec:eight}
we show how to triangulate the knot complement in a case of the
figure-eight knot.
This method can be easily applied to other knots and links.
The last section is devoted to discussions and concluding remarks.




\section{Quantum Dilogarithm Function}
\label{sec:dilog}

\subsection{Classical Dilogarithm Function}

We collect properties concerning the classical dilogarithm function
(see Refs.~\citen{LLewi91Book,Kiril94} for review).
The Euler dilogarithm function $\Li(x)$ is defined by
\begin{align}
  \Li(x)
  =
  \sum_{n=1}^\infty \frac{x^n}{n^2}
  = - \int_0^x
  \frac{\log(1-s)}{s} \, \mathrm{d} s  .
  \label{Euler_integral}
\end{align}
The range $|x| \leq 1$ in an infinite series is extended outside the
unit circle in the second
integral form~\eqref{Euler_integral}.
We later use the Rogers dilogarithm function  defined by
\begin{equation}
  \label{Rogers}
  L(z)
  =
  \Li(z) + \frac{1}{2} \log z \, \log(1-z) .
\end{equation}

Based on the integral form of the dilogarithm function,
we have  the following identities (due to Euler);
\begin{gather}
  \label{Euler_1}
  \Li(z) + \Li(-z) = \frac{1}{2} \, \Li(z^2) ,
  \\[2mm]
  \Li(-z) + \Li(-z^{-1})
  = 2 \, \Li (-1) - \frac{1}{2} \bigl( \log  z  \bigr)^2,
  \label{inverse_Euler}
  \\[2mm]
  \Li(z) + \Li(1-z) = \Li(1) - \log z \, \log(1-z) .
\end{gather}
The first two identities are respectively called the duplication and inversion
relations.
By setting
$z= \mathrm{e}^{\mathrm{i} \, \pi}$ in those identities, we
get
\begin{align}
  \Li(1) &=
  \frac{\pi^2}{6} ,
  &
  \Li(-1) &=
  - \frac{\pi^2}{12} .
\end{align}
Besides above equations, we have a two-variable equation, which we
call  the pentagon identity
 (this form was first written by Schaeffer);
\begin{equation}
  \label{pentagon_Euler}
%
  \Li(\frac{1-x^{-1}}{1-y^{-1}})
  =
  \Li(x) - \Li(y) + \Li(\frac{y}{x})
  +
  \Li(\frac{1-x}{1-y})
  -\frac{\pi^2}{6}
  + \log x \log (\frac{1-x}{1-y}) ,
\end{equation}
or
\begin{equation}
      L(x) - L(y) + L(\frac{y}{x})
    -L(\frac{1-x^{-1}}{1-y^{-1}})
    +L(\frac{1-x}{1-y}) = \frac{\pi^2}{6} .
\end{equation}

It is known that the variant of the dilogarithm function appears in
the 3-dimensional hyperbolic geometry.
Due to Refs.~\citen{WPThurs80Lecture,JMiln82a}, the  volume of the
ideal tetrahedron in
the 3-dimensional hyperbolic space is given by the Bloch--Wigner
function $D(z)$, which is defined by
\begin{equation}
  \label{Bloch_Wigner}
  D(z)
  = \Im \, \Li(z) + \arg(1-z) \cdot \log |z| .
\end{equation}
Here $z$ is a complex parameter $\Im \, z >0$,
which parameterizes the ideal tetrahedron;
the Euclidean triangle cut out of any vertex of the ideal tetrahedron
is similar to that in Fig.~\ref{fig:triangle}.
{}From eq.~\eqref{Euler_1}--\eqref{pentagon_Euler}  we get
\begin{subequations}
  \label{Bloch_relation}
  \begin{gather}
    D(z)
    =
    -D(z^{-1})
    = - D(1-z)
    ,
    \\[2mm]
    D(x) - D(y) + D(\frac{y}{x})
    - D(\frac{1 - x^{-1}}{1 - y^{-1}})
    + D(\frac{1 - x}{1- y})
    = 0 .
  \end{gather}
\end{subequations}
Using the first identity
we can extend a modulus $z$ to  $z\in\mathbb{C} \setminus \{0,1\}$
by regarding
$D(z)$ as the signed
volume of the oriented tetrahedron.

\begin{figure}[htbp]
  \centering
  \begin{equation*}
    \xy/r1.0pc/:.
    {
      { (0,0) \ar@{-} (10,0) },
      { (0,0) \ar@{-} (6,8) },
      { (10,0) \ar@{-} (6,8) },
      { (-0.5,-0.5)*+{0} },
      { (10.5,-0.5)*+{1} },
      { (6,8.5)*+{z} },
      { (1,0.5)*+{z} },
      { (8.2,1)*+{\frac{1}{1-z}} },
      { (5.7,5.8)*+{{1-\frac{1}{z}}} },
      }
    \endxy
  \end{equation*}
  \caption{}
  \label{fig:triangle}
\end{figure}

\subsection{Quantum Dilogarithm Function and the $R$ Operator}

We define a function  $\Phi_\gamma(\varphi)$ by an integral form
following Ref.~\citen{LFadd96a}:
\begin{equation}
  \label{Faddeev_integral}
  \Phi_\gamma(\varphi)
  =
  \exp
  \left(
    \int_{\mathbb{R}+\mathrm{i} \, 0}
    \frac{\mathrm{e}^{- \mathrm{i} \, \varphi \, x}}
    {
      4 \sh ( \gamma \, x) \, \sh(\pi \, x)
      } \,
    \frac{\mathrm{d} x}{x}
  \right) ,
\end{equation}
where we take $\gamma \in \mathbb{R}$.
We note that in Ref.~\citen{Ruijse97a} an essentially same integral
was introduced
in a context of the hyperbolic gamma function, and that
in Ref.~\citen{SLWoron00a} another integral was studied  as the
quantum exponential function which solves the same functional
equations
below.
Also the integral~\eqref{Faddeev_integral} was used to compute the
asymptotic form of Kashaev's invariant~\cite{Kasha96b}.
The function $\Phi_\gamma(\varphi)$
is known as a quantization of the dilogarithm function, and
we have in a limit $\gamma \to 0$
\begin{equation}
  \label{Phi_gamma_0}
  \Phi_\gamma(\varphi)
  \sim
  \exp
  \left(
    \frac{1}{2 \,  \mathrm{i} \, \gamma} \,
    \Li(- \mathrm{e}^\varphi)
  \right) .
%
\end{equation}

We list below several interesting properties of the integral
$\Phi_\gamma(\varphi)$.
\begin{itemize}
\item Duality,
  \begin{equation}
    \Phi_{\frac{\pi^2}{\gamma}} ( \varphi)
    =
    \Phi_{\gamma}(\frac{\gamma}{\pi} \, \varphi) .
  \end{equation}

\item Zero points,
  \begin{equation}
    \text{zeros of $\bigl( \Phi_\gamma(x) \bigr)^{\pm 1}$}
    =
    \Bigl\{
    \mp \mathrm{i} \,
    \bigl(
    (2 \, m +1 ) \, \gamma +
    ( 2 \, n + 1 ) \, \pi
    \bigr)
    \ \big| \
    m, n \in \mathbb{Z}_{\geq 0}
    \Bigr\}
  \end{equation}

\item  Inversion relation,
  \begin{equation}
    \label{inversion_Phi}
    \Phi_\gamma(x) \cdot \Phi_\gamma( -x)
    =
    \exp
    \left(
      - \frac{1}{2 \, \mathrm{i} \, \gamma}
      \Bigl(
      \frac{x^2}{2} + \frac{\pi^2 + \gamma^2}{6}
      \Bigr)
    \right) .
  \end{equation}
  By taking a limit $\gamma \to 0$ and using eq.~\eqref{Phi_gamma_0},
  we obtain
  \begin{equation*}
    \Li(- \mathrm{e}^x) + \Li(-\mathrm{e}^{-x})
    +
    \frac{x^2}{2} + \frac{\pi^2}{6} = 0 ,
  \end{equation*}
  which is nothing but
  the inversion identity~\eqref{inverse_Euler}  for the Euler
  dilogarithm function.

\item Difference equations,
  \begin{subequations}
    \begin{align}
      \frac{\Phi_\gamma(\varphi + \mathrm{i} \, \gamma)}
      {    \Phi_\gamma(\varphi - \mathrm{i} \, \gamma)}
      & =
      \frac{1}{1 + \mathrm{e}^\varphi}         ,
      \\[2mm]
      \frac{ \Phi_\gamma(\varphi + \mathrm{i} \, \pi)}{
        \Phi_\gamma(\varphi - \mathrm{i} \, \pi)}
      & =
      \frac{1}{1 + \mathrm{e}^{\frac{\pi}{\gamma}\varphi}} .
    \end{align}
    
  \end{subequations}

\item Pentagon relation~\cite{FaddKash94,ChekFock99a,LFadd99b},
    \begin{equation}
      \label{pentagon_Phi}
      \Phi_\gamma(\Hat{p}) \, \Phi_\gamma(\Hat{q})
      =
      \Phi_\gamma(\Hat{q}) \, \Phi_\gamma(\Hat{p} + \Hat{q}) \,
      \Phi_\gamma(\Hat{p})  ,
    \end{equation}
    where $\Hat{p}$ and $\Hat{q}$ are the canonically conjugate
    operators satisfying the Heisenberg commutation relation,
    \begin{equation}
      \label{canonical}
      [\Hat{p} ~,~ \Hat{q} ] = - 2 \, \mathrm{i} \, \gamma .  
    \end{equation}
    In this sense, the parameter $\gamma$ in the
    integral~\eqref{Faddeev_integral}  is the Planck constant.

  \item
    The Fourier transformation~\cite{FaddKashVolk00a,PonsoTesch00a},
    \begin{subequations}
      \label{Fourier_transform}
      \begin{gather}
        \frac{1}{\sqrt{4 \, \pi \, \gamma}}
        \int \mathrm{d} y \ \Phi_\gamma(y) \,
        \mathrm{e}^{\frac{1}{2 \, \mathrm{i} \, \gamma} \, x \, y}
        =
        \Phi_\gamma(-x + \mathrm{i} \pi + \mathrm{i} \gamma)
        \,
        \mathrm{e}^{
          \frac{1}{2 \, \mathrm{i} \, \gamma}
          \left(
            \frac{x^2}{2}
            - \frac{1}{2} \pi \gamma
            -
            \frac{\pi^2 + \gamma^2}{6}
          \right)
          } ,
        \\[2mm]
        \frac{1}{\sqrt{4 \, \pi \, \gamma}}
        \int \mathrm{d} y \
        \frac{1}{\Phi_\gamma(y)} \,
        \mathrm{e}^{
          \frac{1}{2 \, \mathrm{i} \, \gamma} x \, y
          }
        =
        \frac{1}{
          \Phi_\gamma(x - \mathrm{i}\, \pi - \mathrm{i}\, \gamma)
          } \,
        \mathrm{e}^{
          -\frac{1}{2 \, \mathrm{i} \, \gamma}
          \left(
            \frac{x^2}{2}
            - \frac{1}{2} \pi \gamma
            - \frac{\pi^2+\gamma^2}{6}
          \right)
          } ,
        \\[2mm]
        \begin{split}
          & \frac{1}{\sqrt{4 \, \pi \, \gamma}}
          \int \mathrm{d} y \
          \frac{\Phi_\gamma(y+u)}{\Phi_\gamma(y+v)} \,
          \mathrm{e}^{
            -\frac{1}{2 \mathrm{i} \gamma} \, x \, y
            }
          \\
          & \qquad =
          \frac{
            \Phi_\gamma(v-u-x+\mathrm{i}\pi + \mathrm{i}\gamma)
            }{
            \Phi_\gamma(v-u+\mathrm{i}\pi + \mathrm{i}\gamma) \,
            \Phi_\gamma(-x-\mathrm{i}\pi - \mathrm{i}\gamma)
            } \,
          \mathrm{e}^{
            \frac{1}{2 \mathrm{i} \gamma}
            \left(
              x ( u- \mathrm{i} \pi - \mathrm{i} \gamma)
              +
              \frac{1}{2} \pi \gamma
              +
              \frac{\pi^2+\gamma^2}{6}
            \right)
            } .
        \end{split}
      \end{gather}
    \end{subequations}
\end{itemize}

For our later convention, we rewrite the pentagon
identity~\eqref{pentagon_Phi}
into a
simple form.
We   define the $S$-operator  on $\mathbf{V}\otimes\mathbf{V}$ by
\begin{equation}
  \label{define_S_operator}
  S_{1,2}
  =
  \mathrm{e}^{
    \frac{1}{2 \,  \mathrm{i}\, \gamma} \,
    \Hat{q}_1 \, \Hat{p}_2
    } \,
  \Phi_\gamma (\Hat{p}_1 + \Hat{q}_2 - \Hat{p}_2) ,
\end{equation}
where the Heisenberg operators $\Hat{p}_a$
and $\Hat{q}_a$ act on the $a$-th  vector
space $\mathbf{V}$.
Then the pentagon identity~\eqref{pentagon_Phi} can be
rewritten as
\begin{equation}
  \label{pentagon_S}
  S_{2,3} \, S_{1,2}
  =
  S_{1,2} \, S_{1,3} \, S_{2,3} ,
\end{equation}
where $S_{a,b}$ acts on the $a$- and $b$-th spaces of
$\mathbf{V}\otimes \mathbf{V} \otimes \mathbf{V}$.
See that
the  operator,
$
  T_{1,2}
  =
  \mathrm{e}^{\frac{1}{2 \, \mathrm{i} \, \gamma } \, \Hat{q}_1 \,
    \Hat{p}_2}  
$,
which is a prefactor of the $S$-operator~\eqref{define_S_operator},
is a simple solution of  eq.~\eqref{pentagon_S}.
We remark  that
the pentagon identity~\eqref{pentagon_S} is a natural consequence  of
the Heisenberg double~\cite{Kasha95b,Kasha96a}, in which the $S$-operator is
given by
\begin{equation*}
  S = \sum_\alpha e_\alpha \otimes e^\alpha .
\end{equation*}
Here $\{ e_\alpha , e^\alpha \}$ is a set of generators satisfying
\begin{gather*}
  e_\alpha \, e_\beta = \sum_\gamma m^\gamma_{\alpha ~ \beta} \, e_\gamma ,
  \\
  e^\alpha \, e^\beta
  = \sum_\gamma  \mu_\gamma^{\alpha ~ \beta} \, e^\gamma ,
  \\
  e_\alpha \, e^\beta =
  \sum_{\gamma, \rho, \sigma}
  m^\beta_{\rho ~ \gamma} \, \mu_\alpha^{\gamma ~ \sigma} \,
  e^\rho \, e_\sigma .
\end{gather*}

\begin{figure}[htbp]
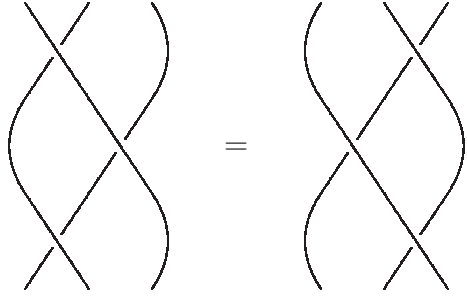

  \centering
  \begin{equation*}
    \xygraph{!{0;/r2pc/:/u3pc/::}="A"
      "A"  [u(.5)]!{\xunderv}
      [urr]!{\xcapv}
      "A"     [d(.5)]!{\xcapv[-1]}
      [ur]!{\xunderv}
      "A"  [d(1.5)] !{\xunderv}
      [urr]!{\xcapv}
      }
    =
    \xygraph{!{0;/r2pc/:/u3pc/::}="A"
      "A"  [u(.5)]!{\xcapv[-1]}
      [ur]!{\xunderv}
      "A"      [d(.5)]!{\xunderv}
      [urr]!{\xcapv}
      "A"     [d(1.5)]!{\xcapv[-1]}
      [ur] !{\xunderv}
      }
    \label{fig:braid_relation}
  \end{equation*}
  \caption{Braid relation, or the constant Yang--Baxter equation.}
  \label{fig:YBE}
\end{figure}

For our purpose to define the knot invariant, we  introduce the
$R$-operator by use of the $S$-operators as~\cite{LFadd99b,Kasha95b,Hikam00a}
\begin{equation}
  \label{R_compact}
  R_{1 2, 3 4}
  =
  \Bigl( S_{1,4}^{\mathrm{t}_4} \Bigr)^{-1} \,
  S_{1,3} \,
  S_{2,4}^{\mathrm{t}_2  \mathrm{t}_4} \,
  \Bigl( S_{2,3}^{\mathrm{t}_2} \Bigr)^{-1} ,
\end{equation}
and we set
\begin{equation}
  \label{check_R}
  \Check{R}_{1 2, 3 4}
  = P_{1,3} \, P_{2,4} \, R_{1 2 , 3 4} .
\end{equation}
Here $\mathrm{t}_a$ means a transposition on the $a$-th space, and $P$
is the permutation operator.
The $R$-operator  acts on a vector space $\mathbf{V}^{\otimes 4}$.
Based on the pentagon identity~\eqref{pentagon_S}, we find that the
$\Check{R}$-operator~\eqref{check_R}
satisfies the Yang--Baxter relation,
\begin{equation}
  \label{YBE_compact}
  \Check{R}_{1 1^\prime, 2 2^\prime} \,
  \Check{R}_{2 2^\prime, 3 3^\prime} \,
  \Check{R}_{1 1^\prime, 2 2^\prime}
  =
  \Check{R}_{2 2^\prime, 3 3^\prime} \,
  \Check{R}_{1 1^\prime, 2 2^\prime} \,
  \Check{R}_{2 2^\prime, 3 3^\prime} .
%
\end{equation}
By regarding the $R$-operator as an operator on
$\mathbf{W} \otimes \mathbf{W}$ with
$\mathbf{W} = \mathbf{V}^{\otimes 2}$,
this Yang--Baxter relation can be seen as a braid relation as usual,
which can be depicted 
as a projection onto 2-dimensional space in Fig.~\ref{fig:YBE}.

\subsection{Representation}
We now give  the  representation of these operators on the momentum
space;
$\Hat{p} \, |p \rangle = p \, | p \rangle$ with $p\in \mathbb{R}$,
and we take the vector space as
$\mathbf{V} = L^2(\mathbb{R})$.
The matrix elements  of the $S$-operators are  given by~\cite{Hikam00a}
\begin{subequations}
  \label{S_element}
\begin{align}
  \langle p_1, p_2 \ | \ S_{1, 2} \ | \ p_1^\prime , p_2^\prime \rangle
  & =
  \frac{1}{4 \, \pi \, \gamma} \,
  \delta(p_1 +  p_2 - p_1^\prime ) \,
  \int \mathrm{d} x \
  \Phi_\gamma(x+p_1) \,
  \mathrm{e}^{
    \frac{1}{2 \, \gamma \, \mathrm{i}} \,
    \left(
      (p_2 - p_2^\prime) \,   x
      -  \frac{1}{2}(p_2 -  p_2^\prime)^2
    \right)
    } ,
  \\[2mm]
  \langle p_1, p_2  \ | \ S_{1, 2}^{~-1} \ | \ p_1^\prime , p_2^\prime \rangle
  & =
  \frac{1}{4 \, \pi \, \gamma} \,
  \delta(p_1 -p_1^\prime - p_2^\prime ) \,
  \int \mathrm{d} x \
  \frac{1}{\Phi_\gamma(x+p_1^\prime)} \,
  \mathrm{e}^{
    \frac{1}{2 \, \gamma \, \mathrm{i}} \,
    \left(
      (p_2 - p_2^\prime) \,   x
      + \frac{1}{2}(p_2 -  p_2^\prime)^2
    \right)
    }  .
\end{align}
\end{subequations}
Due to the Fourier transform~\eqref{Fourier_transform}, these
integrals reduce to
\begin{subequations}
  \label{S_element_2}
  \begin{gather}
    \begin{split}
      & \langle p_1, p_2  \ | \ S_{1, 2}  \ | \ p_1^\prime , p_2^\prime \rangle
      \\
      &  \qquad =
      \frac{1}{\sqrt{4 \, \pi \, \gamma}} \,
      \delta(p_1 + p_2 - p_1^\prime) \cdot
      \Phi_\gamma(p_2^\prime - p_2
      + \mathrm{i} \, \pi + \mathrm{i} \, \gamma) \,
      \mathrm{e}^{\frac{1}{2 \, \mathrm{i} \, \gamma}
        \left(
          - \frac{\pi^2 + \gamma^2}{6} - \frac{\gamma \, \pi}{2}
          + p_1 \, (p_2^\prime - p_2)
        \right)
        } ,
    \end{split}
    \\[4mm]
    \begin{split}
      & \langle p_1, p_2 \ | \ S_{1, 2}^{~-1} \ | \ p_1^\prime , p_2^\prime
      \rangle
      \\
      &  \qquad =
      \frac{1}{\sqrt{4 \, \pi \, \gamma}} \,
      \delta(p_1 - p_1^\prime - p_2^\prime) \,
      \frac{1}{
        \Phi_\gamma(p_2 - p_2^\prime - \mathrm{i} \, \pi
        - \mathrm{i} \, \gamma)
        } \,
      \mathrm{e}^{
        \frac{1}{2 \mathrm{i} \gamma}
        \left(
        \frac{\pi^2 + \gamma^2}{6}
        + \frac{\gamma \, \pi}{2}
        - p_1^\prime ( p_2 - p_2^\prime)
        \right)
        } .
    \end{split}
  \end{gather}
\end{subequations}
The $R$-matrix  is also computed from eqs.~\eqref{R_compact}
and~\eqref{S_element}
as~\cite{Hikam00a}
\begin{subequations}
  \label{R_element}
  \begin{multline}
    \langle p_1, p_2, p_3 , p_4 \ | \  \Check{R}_{1 2, 3 4}  \ | \
    p_1^\prime , p_2^\prime, p_3^\prime, p_4^\prime \rangle
    \\
    =
    \delta(p_1 -p_2+p_3- p_1^\prime) \,
    \delta(p_2^\prime - p_3^\prime - p_4  + p_4^\prime) 
    \\
    \times
    H(p_2^\prime - p_3^\prime , p_3 - p_4^\prime ,
    p_3 - p_2, p_2^\prime  - p_1)  ,
%
  \end{multline}
  \begin{multline}
    \langle p_1, p_2, p_3 , p_4 \  | \ ( \Check{R}_{1 2, 3 4} )^{-1}
    \ | \
    p_1^\prime , p_2^\prime, p_3^\prime, p_4^\prime \rangle
    \\
    =
    \delta(p_4^\prime  -p_2+p_3- p_4) \,
    \delta(p_1 - p_1^\prime - p_3^\prime  + p_2^\prime) 
    \\
    \times
    H(p_4^\prime - p_4, p_1 - p_3 ,
    p_1 - p_1^\prime , p_4^\prime - p_2^\prime)  .
%
  \end{multline}
\end{subequations}
Here the integral $H(a,b,c,d)$ is defined as
\begin{subequations}
  \label{integral_H}
  \begin{align}
    & H(a,b,c,d)
    \nonumber   \\
    & =
    \frac{1}{(4 \, \pi \, \gamma)^2} \,
    \iint \mathrm{d} x \ \mathrm{d} y \
    \frac{
      \Phi_\gamma(x+a) \, \Phi_\gamma(y+c)
      }{
      \Phi_\gamma(x+b) \, \Phi_\gamma(y+d)
      } \,
    \mathrm{e}^{ \frac{1}{2 \,  \mathrm{i} \, \gamma}
      \left(
        -(b-c) \,x + (a-d) \, y - \frac{1}{2}(a-d)^2 - \frac{1}{2}(b-c)^2
      \right)
      }
    \label{integral_H_1}
    \\
    & =
    \frac{1}{4 \, \pi \, \gamma} \,
    \frac{
      \Phi_\gamma(a - b - \mathrm{i} \pi - \mathrm{i} \gamma)
      \cdot
      \Phi_\gamma(d - a + \mathrm{i} \pi + \mathrm{i} \gamma)
      }{
      \Phi_\gamma(c - b - \mathrm{i} \pi - \mathrm{i} \gamma)
      \cdot
      \Phi_\gamma(d - c + \mathrm{i} \pi + \mathrm{i} \gamma)
      }
    \cdot
    \frac{
      \mathrm{e}^{
        \frac{1}{2 \mathrm{i} \gamma} \,
        c( -a+b-c+d)
        }
      }{
      (1- \mathrm{e}^{a-c}) \,
      (1 - \mathrm{e}^{\frac{\pi}{\gamma}(a-c)})
      } ,
  \end{align}
\end{subequations}
where the second equality follows from eq.~\eqref{Fourier_transform}.

As we see that
$H(a,b,c,d) = H(c,d,a,b)$ from eq.~\eqref{integral_H_1}, we have  the
symmetry of the $R$-matrix as
\begin{subequations}
  \label{symmetry_R}
  \begin{gather}
    \langle p_1, p_2 , p_3 , p_4 \ | \  \Check{R} \
    | \  p_1^\prime , p_2^\prime , p_3^\prime , p_4^\prime \rangle
    =
    \langle p_4^\prime , p_3^\prime , p_2^\prime , p_1^\prime \
    | \ \Check{R} \ | \
    p_4 , p_3 , p_2 , p_1 \rangle ,
    \label{R_and_R}
    \\[2mm]
    \langle p_1, p_2 , p_3 , p_4 \  | \ \Check{R}^{-1} \
    | \ p_1^\prime , p_2^\prime , p_3^\prime , p_4^\prime \rangle
    =
    \langle
    p_2^\prime , p_1^\prime, p_1 , p_2 \  | \  \Check{R} \  | \
    p_3^\prime , p_4^\prime , p_4 , p_3 \rangle .
    \label{R_and_R_inv}
  \end{gather}
\end{subequations}

\section{Invariant of Knot and Link}
\label{sec:invariant}

With the ${R}$-matrix
$\Check{R}: \mathbf{W}^{\otimes 2} \to \mathbf{W}^{\otimes 2}$
satisfying the braid
relation~\eqref{YBE_compact},
we can define the invariant of  the  knot $K$.
We assume that we
have the  enhanced Yang--Baxter operators
($\Check{R}, \mu , \alpha, \beta$)
satisfying~\cite{Turae88}
\begin{subequations}
  \label{other_enhanced}
  \begin{gather}
    \bigl( \mu \otimes \mu \bigr) \, \Check{R}
    =
    \Check{R} \, \bigl( \mu \otimes \mu \bigr) ,
    \\[2mm]
    \Tr_2 \Bigl(
    \Check{R}^{\pm 1} \, 
    \bigl( 1 \otimes \mu \bigr)
    \Bigr)
    =
    \alpha^{\pm 1} \, \beta  .
  \end{gather}
\end{subequations}
Here the operator $\mu$ acts on a space $\mathbf{W}$.
When the knot $K$ is given as the closure of
a braid $\xi$ which is represented  in terms of
the Artin $n$ string braid group,
\begin{equation}
  \left\langle
    \sigma_1, \dots, \sigma_{n-1}
    \ \Big| \
    \begin{array}{ll}
      \sigma_i \, \sigma_j = \sigma_j \, \sigma_i,
      &
      \text{for $|i-j| \geq 2$}
      \\[2mm]
      \sigma_i \, \sigma_{i+1} \, \sigma_i
      =
      \sigma_{i+1} \, \sigma_i \, \sigma_{i+1} ,
      &
      \text{for $i=1,\dots,n-1$}
    \end{array}
  \right\rangle ,
\end{equation}
we get the knot invariant $\tau(K)$;
\begin{equation}
  \tau(K)
  =
  \alpha^{- w(\xi)} \, \beta^{-n} \,
  \Tr_{1,\dots, n}
  \Bigl(
  b_R(\xi) \, \mu^{\otimes n}
  \Bigr)  ,
\end{equation}
where $w(\xi)$ is a writhe, \emph{i.e.},  a sum of the exponents, and
$b_R(\xi)$ means   to replace the braid operators $\sigma^{\pm 1}$
by $\Check{R}^{\pm 1}$.
Later we   use another knot  invariant,
\begin{equation}
  \label{tau_1_invariant}
  \tau_1(K)
  =
  \alpha^{- w(\xi)} \,
  \beta^{-n} \,
  \Tr_{2,\dots,n}
  \Bigl(
  b_R(\xi) \, \mu^{\otimes (n-1)}
  \Bigr) ,
\end{equation}
which is associated for ($1,1$)-tangle.

When we use the $\Check{R}$-matrix defined
in eq.~\eqref{R_element},
we find that the $\mu$ operator defined by~\footnote{
The author  thanks
Rinat Kashaev for pointing out an error of
previous manuscript.}
\begin{equation}
  \label{mu_operator}
  \mu =
  \mathrm{e}^{\frac{\pi + \gamma}{\gamma} \, \Hat{p}}
  \otimes
  \mathrm{e}^{ - \frac{\pi + \gamma}{\gamma} \, \Hat{p}^{\mathrm{t}}}
\end{equation}
fulfills eqs.~\eqref{other_enhanced}
with parameters
\begin{align}
  \label{alpha_beta}
  \alpha
  & =
  \mathrm{e}^{- \mathrm{i} \, \frac{\pi^2+\gamma^2}{\gamma}}
  ,
  &
  \beta
  & =
  \frac{
    \gamma \,
    \mathrm{e}^{\mathrm{i} \frac{\pi^2+\gamma^2}{\gamma}}
    }{
    ( 1 - \mathrm{e}^{2 \gamma \mathrm{i}} ) \,
    ( 1 - \mathrm{e}^{2 \frac{\pi^2}{\gamma} \mathrm{i}})
    } .
\end{align}
In this computation  $\Tr$ means integration, and we have used
$
  \delta(x)
  =
  \lim_{\Delta \to 0} \frac{\Delta}{x^2+ \Delta^2} 
$.
As a consequence we have obtained the knot invariant $\tau_1(K)$ from
a set of the
Yang--Baxter operators defined with  eq.~\eqref{R_compact} and
eqs.~\eqref{mu_operator}~--~\eqref{alpha_beta}.
We should stress
that our invariant $\tau_1(K)$
can be viewed as a  non-compact analogue
of Kashaev's invariant which coincides with the colored Jones
polynomial at a specific value as was proved in Ref.~\citen{MuraMura99a}.
In fact
Kashaev's $R$-matrix~\cite{Kasha95} was originally defined based on 
a reduction of
the quantum $S$-operator
when the deformation parameter $\mathrm{e}^{\mathrm{i} \gamma}$
approaches a root of unity.
In  that case the pentagon identity~\eqref{pentagon_S} is
replaced by
\begin{equation*}
    S_{1,2}(p,q,r) \, S_{1,3}(p, q \, r , s) \, S_{2,3}(q,r,s)
  =
  S_{2,3}(p \, q , r , s) \, S_{1,2}(p,q,r \, s)  ,
\end{equation*}
where $p , \dots, s$ are parameters, and a solution $S$  is given in a
finite-dimensional  matrix.

In the following,
we have interests in the asymptotic behavior of the invariant
$\tau_1(K)$, and we define
\begin{equation}
  \label{limit_invariant}
  \Tilde{\tau}_1(K)
  =
  \lim_{\gamma \to 0}
  \Bigl(
  2 \, \mathrm{i} \, \gamma
  \log \tau_1(K)
  \Bigr) .
\end{equation}
As our construction of the invariant is essentially  same with
Kashaev's invariant,
we expect that
the conjecture~\eqref{volume_conjecture} will be applicable to the
invariant~\eqref{limit_invariant}.



\section{Asymptotic Behavior and 3-dimensional Picture}
\label{sec:asymptotic}



\begin{figure}[htbp]
  \begin{center}
    \parbox{2cm}
    {$\langle p_1, p_2 \  | \ S \  | \ p_1^\prime , p_2^\prime \rangle ~:~$}
    \begin{psfrags}
      \psfrag{A}{\textcolor{red}{\reflectbox{$p_1$}}}
      \psfrag{B}{\rotatebox{40}{$p_2$}}
      \psfrag{C}{\rotatebox{-10}{$p_1^\prime$}}
      \psfrag{D}{\textcolor{red}{\reflectbox{$p_2^\prime$}}}
      \psfrag{a}{$z_1$}
      \psfrag{b}{$z_2$}
      \psfrag{c}{$z_3$}
      \psfig{file=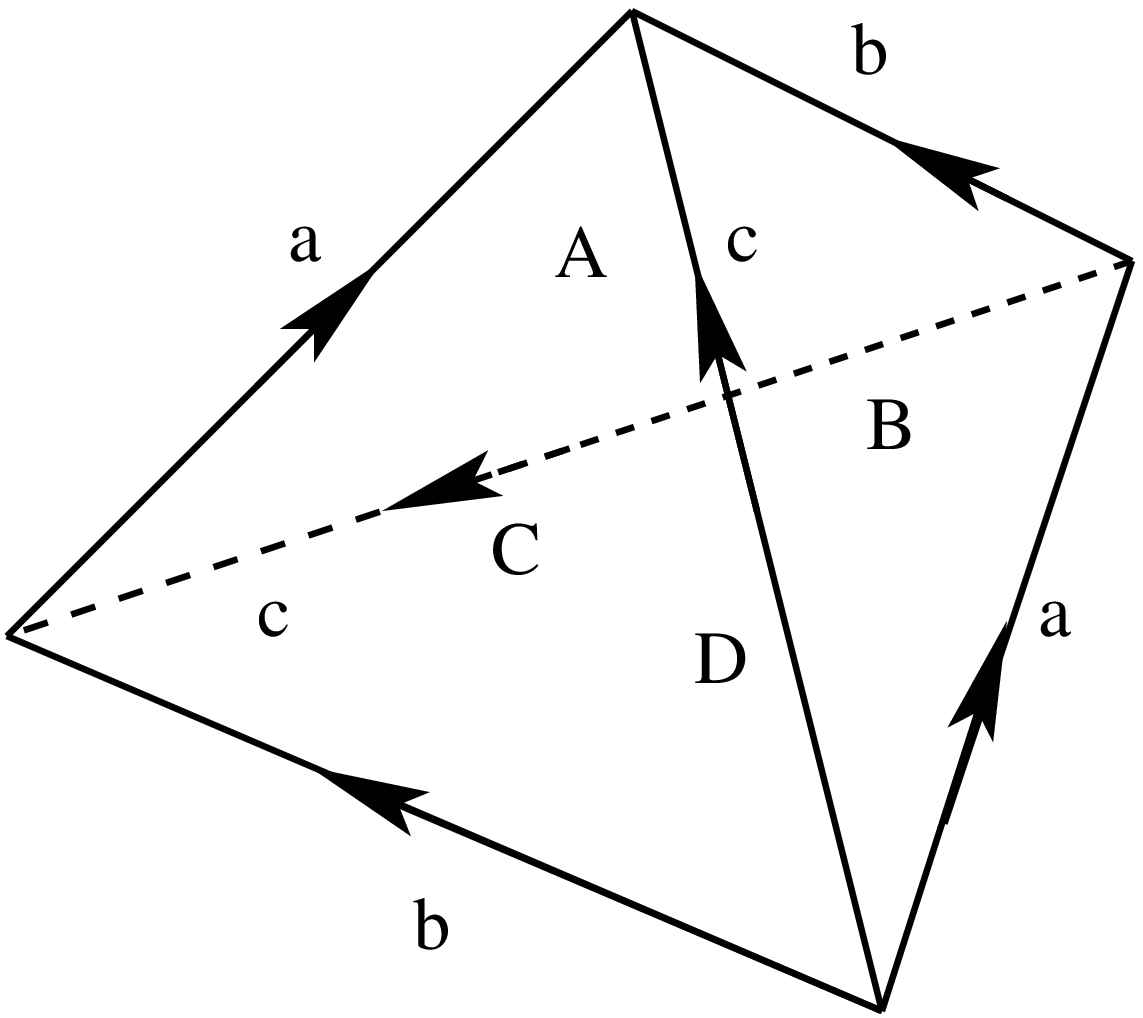,scale=0.4}
    \end{psfrags}
    \hspace{20mm}
    \parbox{2cm}
    {$\langle p_1, p_2 \  | \ S^{-1} \  | \  p_1^\prime , p_2^\prime \rangle ~:~$}
    \begin{psfrags}
      \psfrag{A}{\textcolor{red}{\reflectbox{$p_1$}}}
      \psfrag{B}{\textcolor{red}{\reflectbox{$p_2$}}}
      \psfrag{C}{\rotatebox{-10}{$p_1^\prime$}}
      \psfrag{D}{\rotatebox{40}{$p_2^\prime$}}
      \psfrag{a}{$z_1$}
      \psfrag{b}{$z_2$}
      \psfrag{c}{$z_3$}
      \psfig{file=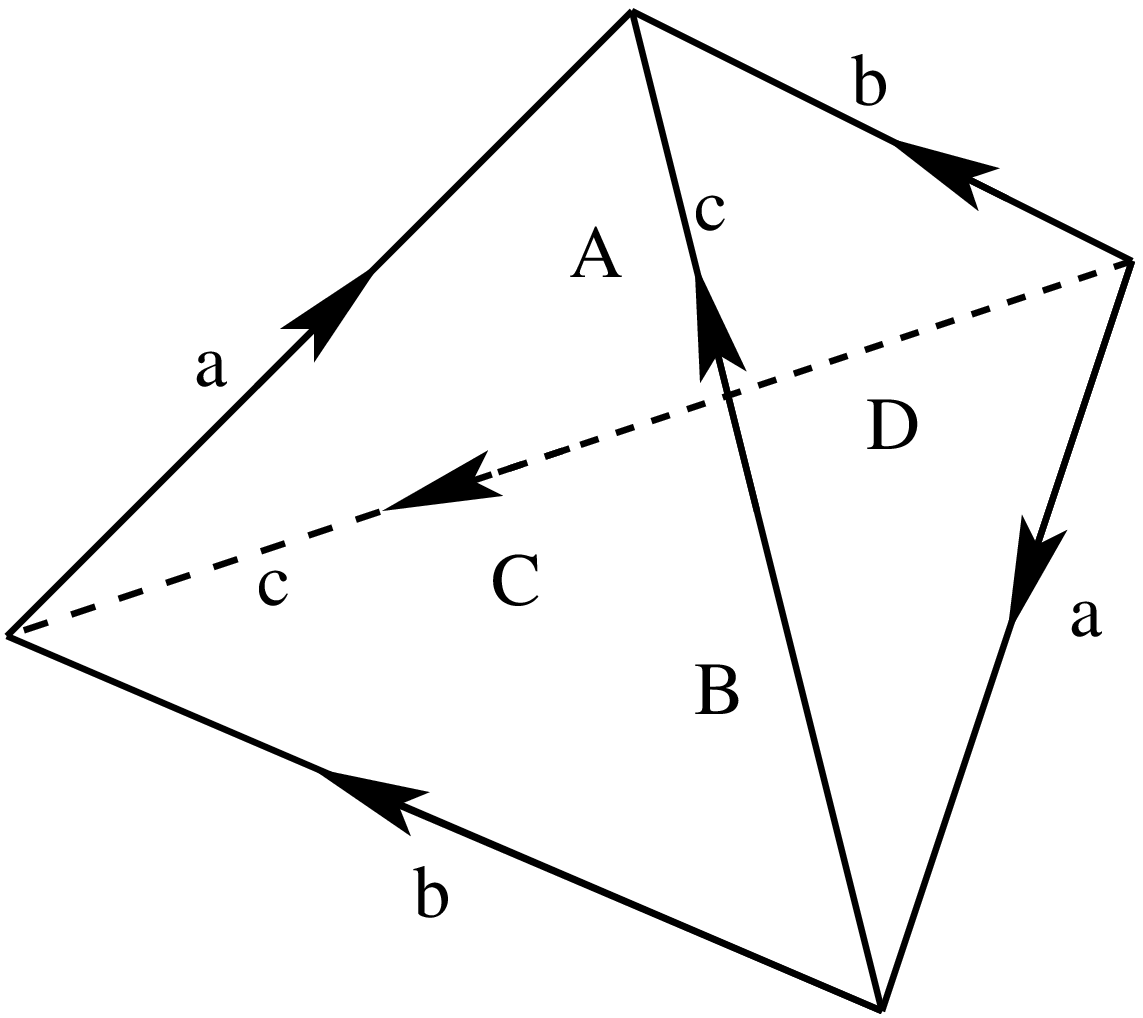,scale=0.4}
    \end{psfrags}
    \caption{Both operators
      $\langle p_1, p_2 \  | \ S \  | \  p_1^\prime , p_2^\prime \rangle$
      and
      $\langle p_1, p_2 \  | \  S^{-1} \  | \  p_1^\prime , p_2^\prime \rangle$
      are represented by the oriented tetrahedra,
      which become the ideal tetrahedra in the classical limit
      $\gamma  \to 0$.
      In this limit they  have  the modulus
      $\mathrm{e}^{p_2^\prime - p_2}$, and
      each edge is associated  by
      $z_1=\mathrm{e}^{p_2^\prime -p_2}$,
      $z_2=1-z_1^{~-1}$, and
      $z_3=(1-z_1)^{-1}$.
      These modulus denote the dihedral angle.
      }
    \label{fig:S_operator}
  \end{center}
\end{figure}

We  shall reveal the 3-dimensional picture  of the knot invariant
$\tau_1(K)$
by studying  an asymptotic behavior in  a limit $\gamma \to 0$.
Using eq.~\eqref{Phi_gamma_0},
we find
that the $S$-operator~\eqref{S_element_2} 
is
represented by
\begin{subequations}
  \label{S_asymptotics}
  \begin{align}
    \langle p_1 , p_2 \ | \ S_{1,2}  \ | \ p_1^\prime , p_2^\prime \rangle
    & \sim
    \delta(p_1 + p_2 - p_1^\prime) \cdot
    \exp
    \left(
      - \frac{1}{2 \, \mathrm{i} \, \gamma}
      \, V(p_2^\prime - p_2 , p_1)
    \right) ,
  \\[2mm]
%
  \langle p_1 , p_2 \  | \ S_{1,2}^{-1} \ | \ p_1^\prime , p_2^\prime \rangle
  &
  \sim
  \delta(p_1 - p_1^\prime - p_2^\prime) \cdot
  \exp
  \left(
    \frac{1}{ 2 \, \mathrm{i} \, \gamma} \,
    V(p_2 - p_2^\prime , p_1^\prime)
  \right) ,
\end{align}
\end{subequations}
where we have defined  the function $V(x,y)$ by
\begin{equation}
  V(x,y)
  =
  \frac{\pi^2}{6} - \Li(\mathrm{e}^x) - x \, y .
\end{equation}
We see that the function $V(x,y)$ is associated with each
$S$-operator, and that
the function $V(x,y)$ has an interesting property for our purpose to
relate with the 3-dimensional hyperbolic geometry;
when we suppose an  analytic continuation  $x,y \in \mathbb{C}$,
we have
\begin{subequations}
  \label{property_V}
  \begin{align}
    V(x,y)
    & =
    L(1- \mathrm{e}^x)
    +
    \frac{1}{2}
    \left(
      x \, \frac{\partial V(x,y)}{\partial x}
      +
      y \, \frac{\partial V(x,y)}{\partial y}
    \right) ,
    \\[2mm]
    \Im \, V(x,y)
    & =
    D(1-\mathrm{e}^x)
    +
    \log | \mathrm{e}^x | \cdot
    \Im \, \biggl( \frac{\partial}{\partial x} V(x,y) \biggr) 
    +
    \log | \mathrm{e}^y | \cdot
    \Im \, \biggl( \frac{\partial}{\partial y} V(x,y) \biggr) .
  \end{align}
\end{subequations}
Here
$L(z)$ is  the Rogers dilogarithm~\eqref{Rogers}, and
the function $D(z)$
is the Bloch--Wigner
function~\eqref{Bloch_Wigner} which gives the hyperbolic volume of
the ideal tetrahedron.
This indicates that  the classical limit of
the $S$-operator
describes
the
ideal tetrahedron at the critical point.
By these observations we can associate the ideal tetrahedron for the
$S$-operators
$\langle p_1, p_2 \ | \ S^{\pm 1} \  | \ p_1^\prime , p_2^\prime \rangle$
as Fig.~\ref{fig:S_operator}.
Due to sign   of the function $V(x,y)$ in eqs.~\eqref{S_asymptotics}
these tetrahedra are mirror images each other.
Therein the modulus of the ideal tetrahedron is
given by $\mathrm{e}^{p_2^\prime - p_2}$ and each face has a momentum;
we regard $p_i$ and $p_i^\prime$ as the outgoing and incoming states
respectively,
\emph{i.e.},
each triangular face
is assigned a transverse orientation.
See that the dihedral angles of opposite edges are equal, and we have
\begin{align*}
  & z_1 \, z_2 \, z_3 = -1 ,
  &
  & 1 - z_1 + z_1 \, z_2 = 0 .
\end{align*}

\begin{figure}[htbp]
  \begin{center}
    \psfig{file=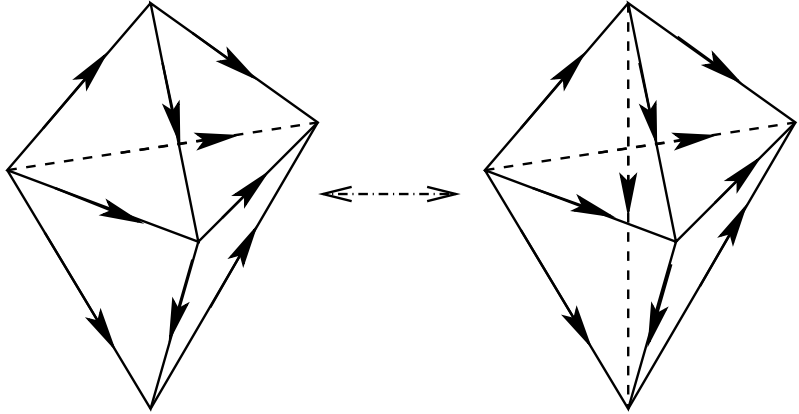,scale=0.5}
    \caption{Pachner move $2 \leftrightarrow 3$.}
    \label{fig:pentagon}
  \end{center}
\end{figure}

Our  identification
of the modulus and dihedral angles
can be justified  from the 3-dimensional picture of the
pentagon identity as follows.
The pentagon identity~\eqref{pentagon_S}  is depicted as
the 2-3 Pachner move
(Fig.~\ref{fig:pentagon})
once we represent the $S$-operators by the oriented tetrahedra as in
Fig.~\ref{fig:S_operator}.

The matrix element of the right hand side of eq.~\eqref{pentagon_S}
is written as
\begin{equation}
  \label{pentagon_rhs}
  \iiint \mathrm{d} y \ \mathrm{d} z \
  \mathrm{d} w \
  \langle p_1, p_2 \ | \  S \  | \ y,z \rangle \,
  \langle y, p_3 \ | \  S \  | \  p_1^\prime ,w \rangle \,
  \langle z, w \ | \ S \  | \  p_2^\prime , p_3^\prime \rangle   .
\end{equation}
After  substituting  the asymptotic form~\eqref{S_asymptotics} into this
integral,
we get immediately
\begin{align*}
  y &= p_1 + p_2 ,
  &
  w &= p_2^\prime -z ,
\end{align*}
and the integral reduces to
\begin{multline*}
  \delta(p_1 + p_2 + p_3 - p_1^\prime) \cdot
  \int \mathrm{d} z \
  \exp \frac{1}{2 \, \mathrm{i} \, \gamma}
  \Bigl(
  -\frac{\pi^2}{2}
  + \Li(\mathrm{e}^{z-p_2})
  + \Li(\mathrm{e}^{p_2^\prime - p_3 - z})
  + \Li(\mathrm{e}^{p_3^\prime - p_2^\prime + z})
  \\
  + z \, (-p_2 + p_3^\prime - p_2^\prime + z)
  - p_1 \, p_2 + ( p_2^\prime - p_3) \, ( p_1 + p_2)
  \Bigr).
\end{multline*}
As we study a limit $\gamma\to 0$, the integral is evaluated by the
saddle point method, whose
saddle point condition  is given by
\begin{equation}
  \label{pentagon_condition}
  ( 1 - \mathrm{e}^{w-p_3} )^{-1} \cdot
  ( 1 - \mathrm{e}^{p_2-z}) \cdot
  ( 1 - \mathrm{e}^{w-p_3^\prime}) = 1 .
\end{equation}
This condition  exactly coincides with the hyperbolicity equation
around an axis which penetrates 2 adjacent 
tetrahedra in the right hand side of
Fig.~\ref{fig:pentagon},
once we regard the $S$-operator as the ideal tetrahedron
whose dihedral angles are written in Fig.~\ref{fig:S_operator}.
See that by substituting  a  solution of
eq.~\eqref{pentagon_condition}
into eq.~\eqref{pentagon_rhs} we  recover the left hand side of
eq.~\eqref{pentagon_S} after  using  Schaeffer's pentagon
identity~\eqref{pentagon_Euler}.

This coincidence between the   saddle point equation and the
hyperbolicity consistency condition can be seen for other pentagon identities,
such as
$S_{2,3} \, S_{1,2} \, S_{2,3}^{~-1} = S_{1,2} \, S_{1,3}$,
$S_{1,2}^{~-1} \, S_{2,3} = S_{1,3} \, S_{2,3} \, S_{1,2}^{~-1}$,
and the trivial identities
$S_{1,2} \, S_{1,2}^{~-1} = 1$.
This fact supports a validity
that at the critical point the classical limit of the  $S$-operator
represents the (oriented) ideal tetrahedron
whose dihedral angles are fixed as in Fig.~\ref{fig:S_operator}.

\section{3-Dimensional Picture of the  Invariant}

\subsection{$R$-Operators as the Braid Operator}

\begin{figure}[htbp]
  \begin{center}
    \begin{psfrags}
      \psfrag{A}{$p_3^\prime$}
      \psfrag{B}{\rotatebox{50}{$p_4^\prime$}}
      \psfrag{C}{\rotatebox{-50}{$p_1$}}
      \psfrag{D}{\textcolor{red}{\reflectbox{$p_2$}}}
      \psfrag{E}{\textcolor{red}{\rotatebox{40}{\reflectbox{$p_1^\prime$}}}}
      \psfrag{F}{$p_2^\prime$}
      \psfrag{G}{\textcolor{red}{\reflectbox{$p_3$}}}
      \psfrag{H}{\textcolor{red}{\rotatebox{-45}{\reflectbox{$p_4$}}}}
      \psfig{file=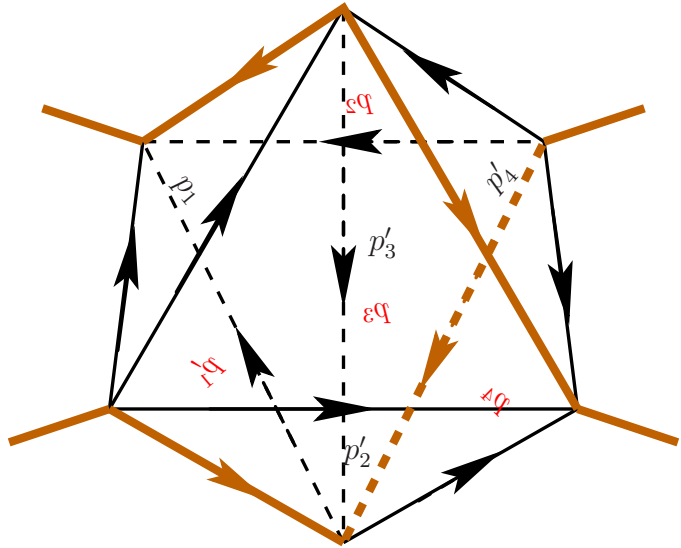,scale=0.7}
    \end{psfrags}
    \caption{The braid operator ($\Check{R}$-matrix)
      $\langle \vec{p}  \ | \  \Check{R}  \ | \ \vec{p^\prime} \rangle$
      is represented by the octahedron.}
    \label{fig:R_matrix}
  \end{center}
\end{figure}

\begin{figure}[htbp]
  \begin{center}
    \begin{psfrags}
      \psfrag{A}{$p_4$}
      \psfrag{B}{\rotatebox{50}{$p_3$}}
      \psfrag{C}{\rotatebox{-50}{$p_2^\prime$}}
      \psfrag{D}{\textcolor{red}{\reflectbox{$p_1^\prime$}}}
      \psfrag{E}{\textcolor{red}{\rotatebox{40}{\reflectbox{$p_3^\prime$}}}}
      \psfrag{F}{$p_4^\prime$}
      \psfrag{G}{\textcolor{red}{\reflectbox{$p_1$}}}
      \psfrag{H}{\textcolor{red}{\rotatebox{-45}{\reflectbox{$p_2$}}}}
      \psfig{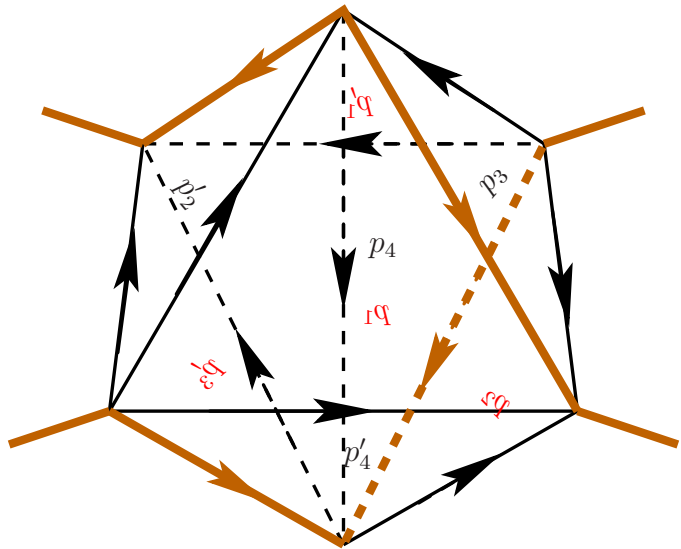}
    \end{psfrags}
    \caption{The inverse of the  $\Check{R}$-matrix,
      $\langle \vec{p} \  | \  \Check{R}^{-1} \ | \  \vec{p^\prime} \rangle$,
      is represented by the same  octahedron with the $\Check{R}$
      operator, though the content is different.}
    \label{fig:R_inverse}
  \end{center}
\end{figure}

We now give the 3-dimensional picture of the
$R$-operators~\eqref{check_R}, and study  the relationship between the
hyperbolic geometry and our invariant.
The $R$-operator consists from four  $S$ operators, and its
matrix element
$\langle \vec{p} \ | \ \Check{R}_{12,34} \ | \  \vec{p^\prime} \rangle$
is explicitly  given by
\begin{multline}
  \label{R_element_int}
  \langle p_1, p_2, p_3, p_4 \  | \
  \Check{R}_{12,34}  \ | \
  p_1^\prime , p_2^\prime, p_3^\prime, p_4^\prime \rangle
  \\
  =
  \iiiint \mathrm{d} x \ \mathrm{d} y \ \mathrm{d} z \
  \mathrm{d} w \
  \langle p_3, w \ | \  S^{-1}  \ | \  x, p_2 \rangle \,
  \langle y, p_4^\prime \ | \  S \  | \  p_4 , w \rangle \,
  \langle p_2^\prime,  z \ | \  S^{-1} \  | \  y, p_3^\prime \rangle \,
  \langle x, p_1 \  | \  S \  | \  p_1^\prime , z \rangle   .
\end{multline}
This integration can be performed explicitly  as eq.~\eqref{integral_H}, but we
work with this form to see a gluing condition clearly.
As we regard the  $S$-operators as the oriented (ideal) tetrahedron
(Fig.~\ref{fig:S_operator}),
the $\Check{R}$-operator is depicted as the oriented octahedron  in
Fig.~\ref{fig:R_matrix} by gluing faces of tetrahedra  to each other.
{}From the symmetry of the $R$-matrix~\eqref{R_and_R_inv}, the
operator $\Check{R}^{-1}$ is  written as the
octahedron  in Fig.~\ref{fig:R_inverse}.
Assignment of the octahedron to the braiding operator
first appeared  in Ref.~\citen{DThurs99a}, and
it was   later
used to give the  decomposition of the knot complement
directly from  Kashaev's invariant~\cite{YYokot00b}.
Our result in  Fig.~\ref{fig:R_matrix} is essentially same with one in
Ref.~\citen{DThurs99a}, and  this
agreement  suggests  that
our knot invariant $\tau_1(K)$  is indeed  defined as
a non-compact analogue of
Kashaev's invariant (the colored Jones polynomial at a  fixed value),
only replacing the
\emph{finite}-dimensional representation of the quantum dilogarithm
function with the \emph{infinite}-dimensional one.
Consequently  the decomposition of the knot complement which will be presented
below
is same with one
given in Ref.~\citen{YYokot00b}
(see also Ref.~\citen{HMuraka00c}).

The braiding property can be seen from
the realization of the $R$-operators as in
Figs.~\ref{fig:R_matrix}--\ref{fig:R_inverse}.
When we suppose that the gray bold lines in those figures
denote the link $L$ and we
look down each octahedron from the top,
we find that the braiding is indeed  realized as follows;
\begin{subequations}
  \label{realize_R}
  \begin{align}
    \langle \vec{p} \ | \ \Check{R} \ | \ \vec{p^\prime} \rangle
    & = \quad
    \xygraph{!{0;/r4pc/:/u4pc/::}="A"
      "A"  [d(0.5)]!{\xunderv}
         !{(-0.09,.4)*+{p_1}}
         !{(0.09,.6)*+{p_2}}
         !{(0.91,.6)*+{p_3}}
         !{(1.1,.38)*+{p_4}}
         !{(-0.09,-0.4)*+{p_1^\prime}}
         !{(0.09,-0.6)*+{p_2^\prime}}
         !{(0.91,-0.6)*+{p_3^\prime}}
         !{(1.1,-0.38)*+{p_4^\prime}}
         }
\\[2mm]
    \langle \vec{p} \ | \  \Check{R}^{-1}  \ | \ \vec{p^\prime} \rangle
    & = \quad
    \xygraph{!{0;/r4pc/:/u4pc/::}="A"
      "A"  [d(0.5)]!{\xoverv}
         !{(-0.09,.4)*+{p_1}}
         !{(0.09,.6)*+{p_2}}
         !{(0.91,.6)*+{p_3}}
         !{(1.1,.38)*+{p_4}}
         !{(-0.09,-0.4)*+{p_1^\prime}}
         !{(0.09,-0.6)*+{p_2^\prime}}
         !{(0.91,-0.6)*+{p_3^\prime}}
         !{(1.1,-0.38)*+{p_4^\prime}}
         }
     \end{align}
   \end{subequations}
We should stress  that  every 0-simplex  of the octahedron is on the
link $L$, and that the octahedron is in the complement of the link $L$.

We shall check the hyperbolicity consistency  condition in constructing
the  $R$-operator from
the ideal tetrahedra in a case of $\gamma \to 0$.
Substituting the asymptotic form~\eqref{S_asymptotics}
into
eq.~\eqref{R_element_int},
we get
\begin{multline}
  \langle \vec{p} \ | \  \Check{R}_{12,34} \  | \ \vec{p^\prime} \rangle
  \sim
  \delta(p_1 - p_2 + p_3 - p_1^\prime) \,
  \delta(p_2^\prime -p_3^\prime + p_4^\prime - p_4) \\
  \times
  \iint \mathrm{d} z \ \mathrm{d} w \
  \exp
  \frac{1}{ 2 \, \mathrm{i} \, \gamma}
  \biggl(
  - \Li(\mathrm{e}^{w-p_2}) + \Li(\mathrm{e}^{w-p_4^\prime})
  - \Li(\mathrm{e}^{z-p_3^\prime})
  + \Li(\mathrm{e}^{z-p_1})
  \\
  + (-p_1 + p_1^\prime) \, (-w+p_2 + z - p_1)
  + (p_4 - p_4^\prime) \, ( w - p_4^\prime - z + p_3^\prime)
  \biggr) .
\end{multline}
Here we have used  trivial  constraints;
\begin{align}
  \label{constraint_x_y}
  x & = -p_1 + p_1^\prime  ,
  &
  y &=  p_4 - p_4^\prime .
\end{align}
Above  integral is  evaluated by the saddle point,
in which
we have constraints
  \begin{align*}
    & \frac{1- \mathrm{e}^{w-p_2}}{1-\mathrm{e}^{w-p_4^\prime}}
    \,
    \mathrm{e}^{p_1 - p_1^\prime + p_4 - p_4^\prime}
    = 1 ,
    &
    &
    \frac{1- \mathrm{e}^{z-p_3^\prime}}{1-\mathrm{e}^{z-p_1}}
    \,
    \mathrm{e}^{-p_1 + p_1^\prime - p_4 + p_4^\prime}
    = 1  .
  \end{align*}
This set of equations solves
\begin{align}
  \label{obtain_w_z}
  \mathrm{e}^w
  &=
  \frac{1- \mathrm{e}^{p_1 - p_1^\prime + p_4 - p_4^\prime}}
  {
    \mathrm{e}^{-p_4^\prime}
    -
    \mathrm{e}^{p_1 - p_1^\prime + p_4 - p_4^\prime - p_2}
    } ,
  &
  \mathrm{e}^z
  &=
  \frac{1- \mathrm{e}^{p_1 - p_1^\prime + p_4 - p_4^\prime}}
  {
    \mathrm{e}^{-p_3^\prime}
    -
    \mathrm{e}^{p_1 - p_1^\prime + p_4 - p_4^\prime - p_1}
    } ,
\end{align}
and we easily find a constraint,
\begin{equation}
  \label{constraint_w_z}
  \frac{1- \mathrm{e}^{w-p_2}}{1-\mathrm{e}^{w-p_4^\prime}} \cdot
  \frac{1- \mathrm{e}^{z-p_3^\prime}}{1-\mathrm{e}^{z-p_1}}
  = 1 ,
\end{equation}
which
coincides with  the hyperbolicity condition around vertical axis
(crossing point in eq~\eqref{realize_R}).

Due to the symmetry of the $R$-operator~\eqref{symmetry_R},
we can conclude that
for each crossing in a link  $L$
we can attach the oriented
octahedron, which has a following projection;
\begin{equation}
  \label{attach_octa}
  \xy/r1.2pc/:.
  {
    { (0,0)*+{\text{\LARGE$\bigotimes$}} },
    { (0,0) \ar@{-}|@{>} (4,-4)},
    { (0,0) \ar@{-}|@{>} (-4,4)},
    { (1.5,1.5) \ar@{-}|@{<} (4,4)},
    { (-1.5,-1.5) \ar@{-}|@{<} (-4,-4)},
    { (4,4) \ar@{--}|@{>} (4,-4)},
    { (4,4) \ar@{--}|@{>} (-4,4)},
    { (-4,-4) \ar@{--}|@{>} (4,-4)},
    { (-4,-4) \ar@{--}|@{>} (-4,4)},
    { (4,4) \ar@{-}|@{} (6,6)},
    { (4,-4) \ar@{-}|@{} (6,-6)},
    { (-4,4) \ar@{-}|@{} (-6,6)},
    { (-4,-4) \ar@{-}|@{} (-6,-6)},
    { (-6,5)*+{p_1} },
    { (-5,6)*+{p_2} },
    { (5,6)*+{p_3} },
    { (6,5)*+{p_4} },
    { (-6,-5)*+{p_1^\prime} },
    { (-5,-6)*+{p_2^\prime} },
    { (5,-6)*+{p_3^\prime} },
    { (6,-5)*+{p_4^\prime} },
    { (0,5)*+{\langle p_3, w \ | \  S^{-1}  \ | \  x, p_2 \rangle} },
    { (0,-5)*+{\langle p_2^\prime, z \ | \  S^{-1} \  | \  y, p_3^\prime \rangle} },
    { (-5,0)*+{
        \text{
          \rotatebox{-90}{
            $\langle x, p_1 \ | \  S  \ | \  p_1^\prime, z \rangle$}
        }}
      },
    { (5,0)*+{
        \text{
          \rotatebox{90}{
            $\langle y, p_4^\prime \ | \  S \  | \  p_4, w \rangle$}
        }}
      },
%
    { (2,2.5)*+{
        \text{
          \rotatebox{45}{
            \textcolor{red}{$w$}
            } }} },
    { (-2,-2.5)*+{
        \text{
          \rotatebox{45}{
            \textcolor{red}{$z$}
            } }} },
    { (-2,1.5)*+{
        \text{
          \rotatebox{-45}{
            \textcolor{red}{$x$}
            } }} },
    { (2,-1.5)*+{
        \text{
          \rotatebox{-45}{
            \textcolor{red}{$y$}
            } }} },
        { (0,1.1)*+{
        \text{
          $a_4$
              } }},
        { (1.1,0)*+{
        \text{
          \rotatebox{90}{
            $a_3$
            } }} },
    { (-1.1,0)*+{
        \text{
          \rotatebox{-90}{
            $a_1$
          }} }},
    { (0,-1.1)*+{
        \text{
          $a_2$
          }} },
    }
  \endxy
\end{equation}
Here $x$ and $y$ are auxiliary momenta
given by eq.~\eqref{constraint_x_y}, and $w$ and $z$
are fixed by eq.~\eqref{obtain_w_z}.
We further have
\begin{align}
  p_3
  & = - p_1 + p_2 + p_1^\prime ,
  &
  p_2^\prime
  & = p_4 + p_3^\prime - p_4^\prime  .
\end{align}
and the dihedral angles $a_i$ satisfying
$a_1 \, a_2 \, a_3 \, a_4 = 1$ are given by
\begin{align*}
  a_1
  &=
  (1- \mathrm{e}^{z-p_1})^{-1}
  =
  \frac{
    \mathrm{e}^{p_1 - p_1^\prime + p_4 - p_4^\prime}
    - \mathrm{e}^{p_1 - p_3^\prime}}
  {1 - \mathrm{e}^{p_1 - p_3^\prime}} ,
  \\
  a_2
  &=
  1- \mathrm{e}^{z-p_3^\prime}
  =
  \frac{1 - \mathrm{e}^{p_1 - p_3^\prime}}
  {
    1
    - \mathrm{e}^{-p_1 + p_1^\prime - p_4 + p_4^\prime}
    \mathrm{e}^{p_1 - p_3^\prime}
    } ,
  \\
  a_3
  &=
  (1 - \mathrm{e}^{w-p_4^\prime})^{-1}
  =
  \frac{
    1 
    - \mathrm{e}^{-p_1 + p_1^\prime - p_4 + p_4^\prime}
    \mathrm{e}^{p_2 - p_4^\prime}
    } 
  {1 - \mathrm{e}^{p_2 - p_4^\prime}} ,
  \\
  a_4
  &=
  1 - \mathrm{e}^{w-p_2}
  =
  \frac{1 - \mathrm{e}^{p_2 - p_4^\prime}}
  {
    \mathrm{e}^{p_1 - p_1^\prime + p_4 - p_4^\prime}
    -
    \mathrm{e}^{p_2 - p_4^\prime}
    } .
\end{align*}

To close this section,
we give an explicit form of an asymptotic form of the
$\Check{R}$-operators.
As
the integral~\eqref{integral_H} has an asymptotic
form,
\begin{multline}
  H(a,b,c,d) \\
  \sim
  \exp \frac{1}{2 \, \mathrm{i} \, \gamma}
  \biggl(
  \Li(\mathrm{e}^{a-b}) + \Li(\mathrm{e}^{d-a})
  - \Li(\mathrm{e}^{c-b}) - \Li(\mathrm{e}^{d-c})
  + c\, ( -a + b - c + d)
  \biggr) ,
\end{multline}
the $\Check{R}^{\pm 1}$ operators  in a limit
$\gamma \to 0$
are
respectively given by
\begin{subequations}
\label{R_asymptotics}
\begin{multline}
  \langle \vec{p} \ | \
  \Check{R}_{1 2, 3 4} \
  | \ \vec{p^\prime} \rangle
  \sim
  \delta(p_1 + p_3 - p_2 - p_1^\prime ) \cdot
  \delta(p_2^\prime - p_3^\prime + p_4^\prime - p_4)
  \\
  \times
  \exp
  \frac{1}{2 \, \mathrm{i} \, \gamma}
  \biggl(
  \Li( \mathrm{e}^{p_4  - p_3} )
  +  \Li( \mathrm{e}^{p_3^\prime - p_1} )
  -  \Li( \mathrm{e}^{p_4^\prime - p_2} )
  -  \Li( \mathrm{e}^{p_2^\prime - p_1^\prime} )
  \\
  + ( p_1^\prime - p_1) \,
  (-p_4 + p_3 - p_1^\prime + p_2^\prime)
  \biggr) ,
%
\end{multline}
\begin{multline}
  \langle \vec{p} \ |  \
  ( \Check{R}_{1 2, 3 4} )^{-1} \ | \
  \vec{p^\prime} \rangle
  \sim
  \delta(p_4^\prime  -p_2+p_3- p_4) \,
  \delta(p_1 - p_1^\prime - p_3^\prime  + p_2^\prime) 
  \\
  \times
  \exp \frac{1}{2 \, \mathrm{i} \, \gamma}
  \biggl(
    \Li(\mathrm{e}^{p_2 - p_1}) 
    + \Li(\mathrm{e}^{p_4 - p_2^\prime})
    - \Li(\mathrm{e}^{p_3 - p_1^\prime})
    - \Li(\mathrm{e}^{p_4^\prime - p_3^\prime})
    \\
    + (p_1 - p_1^\prime) \, (p_1 - p_2 - p_3^\prime + p_4^\prime)
  \biggr) .
\end{multline}
\end{subequations}

In constructing the knot invariant $\tau_1(K)$, we need another
operator $\mu$~\eqref{mu_operator}.
The  matrix element of the  $\mu$-operator
can be computed simply,
and  in the classical  limit $\gamma \to 0$ reduces to
\begin{equation}
  \langle p_1 , p_2 \ | \ \mu \ | \ p_1^\prime , p_2^\prime \rangle
  \sim
  \delta(p_1 - p_1^\prime) \, \delta(p_2 - p_2^\prime) \cdot
  \exp \frac{\pi}{\gamma} \bigl( p_1 - p_2 \bigr) .
\end{equation}
Thus
the saddle point condition coming  from the $\mu$-operator is always
$2 \, \pi \, \mathrm{i}$, and
we can ignore
a  contribution  from the $\mu$-operator
to the saddle point condition.

In the rest of this   section, we 
study how to glue these octahedra in the invariant $\tau_1(K)$.
We   show that for every gluing
there exists  a  correspondence between the
saddle point equations and the hyperbolicity consistency conditions
around edge.

\subsection{Hyperbolicity Condition for Surface}
\begin{figure}[t]
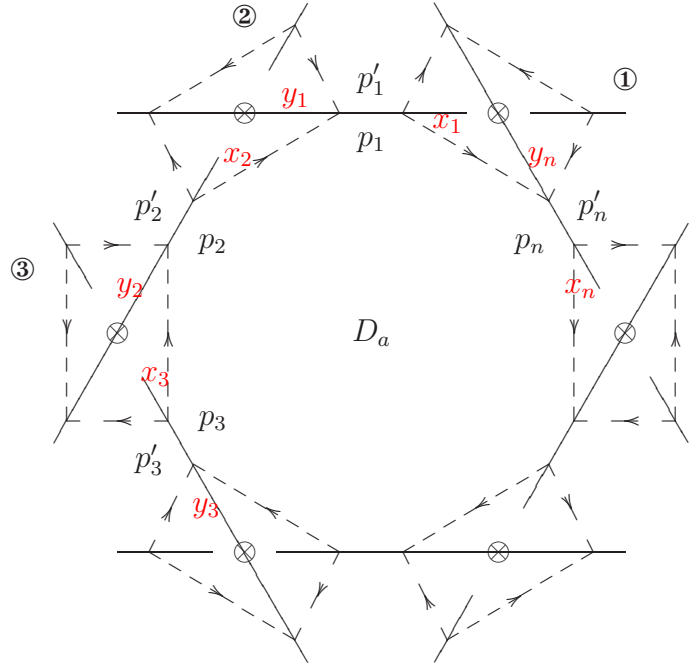

  \begin{equation*}
    \xy/r2.0pc/:.
    {
      { (0,0)*+{D_a} },
      { (5, -1.73) \ar@{-}|@{} (4.4, -0.7) },
      { (3.6, 0.7) \ar@{-}|@{} (1, 5.2) },
      { (4, 3.46) \ar@{-}|@{} (2.5, 3.46) },
      { (1.5, 3.46) \ar@{-}|@{} (-4, 3.46) },
      { (-1, 5.2) \ar@{-}|@{} (-1.6, 4.15) },
      { (-2.4, 2.77) \ar@{-}|@{} (-5, -1.73) },
      { (-5, 1.73) \ar@{-}|@{} (-4.4, 0.7) },
      { (-3.6, -0.7) \ar@{-}|@{} (-1, -5.2) },
      { (-4, -3.46) \ar@{-}|@{} (-2.5, -3.46) },
      { (-1.5, -3.46) \ar@{-}|@{} (4, -3.46) },
      { (1, -5.2) \ar@{-}|@{} (1.6, -4.15) },
      { (2.4, -2.77) \ar@{-}|@{} (5, 1.73) },
      { (4,0)*+{\otimes} },
      { (2, 3.46)*+{\otimes} },
      { (-2, 3.46)*+{\otimes} },
      { (-4,0)*+{\otimes} },
      { (2, -3.46)*+{\otimes} },
      { (-2, -3.46)*+{\otimes} },
      { (2.8, 2.08) \ar@{--}|@{<} (3.5, 3.46)  },
      { (2.8, 2.08) \ar@{--}|@{<} (0.5, 3.46)  },
      { (3.5, 3.46)  \ar@{--}|@{>} (1.2, 4.85) },
      { (0.5, 3.46)  \ar@{--}|@{>} (1.2, 4.85) },
      { (3.2, 1.39) \ar@{--}|@{>} (4.8,1.39)  },
      { (3.2, 1.39) \ar@{--}|@{>} (3.2, -1.39)  },
      { (4.8, -1.39) \ar@{--}|@{>} (4.8,1.39)  },
      { (4.8, -1.39) \ar@{--}|@{>} (3.2, -1.39)  },
      { (-3.2, 1.39) \ar@{--}|@{<} (-4.8,1.39)  },
      { (-3.2, 1.39) \ar@{--}|@{<} (-3.2, -1.39)  },
      { (-4.8, -1.39) \ar@{--}|@{<} (-4.8,1.39)  },
      { (-4.8, -1.39) \ar@{--}|@{<} (-3.2, -1.39)  },
      { (-2.8, 2.08) \ar@{--}|@{>} (-3.5, 3.46)  },
      { (-2.8, 2.08) \ar@{--}|@{>} (-0.5, 3.46)  },
      { (-3.5, 3.46)  \ar@{--}|@{<} (-1.2, 4.85) },
      { (-0.5, 3.46)  \ar@{--}|@{<} (-1.2, 4.85) },
      { (-2.8, -2.08) \ar@{--}|@{<} (-3.5, -3.46)  },
      { (-2.8, -2.08) \ar@{--}|@{<} (-0.5, -3.46)  },
      { (-3.5, -3.46)  \ar@{--}|@{>} (-1.2, -4.85) },
      { (-0.5, -3.46)  \ar@{--}|@{>} (-1.2, -4.85) },
      { (2.8, -2.08) \ar@{--}|@{>} (3.5, -3.46)  },
      { (2.8, -2.08) \ar@{--}|@{>} (0.5, -3.46)  },
      { (3.5, -3.46)  \ar@{--}|@{<} (1.2, -4.85) },
      { (0.5, -3.46)  \ar@{--}|@{<} (1.2, -4.85) },
%
      { (0,4)*+{p_1^\prime} },
      { (0,3)*+{p_1} },
      { (-3.5,2)*+{p_2^\prime} },
      { (-2.5,1.4)*+{p_2} },
      { (-3.5, -2)*+{p_3^\prime} },
      { (-2.5, -1.4)*+{p_3} },
      { (3.5, 2)*+{p_n^\prime} },
      { (2.5, 1.4)*+{p_n} },
      { (1.2, 3.3)*+{\textcolor{red}{x_1}} },
      { (-1.2, 3.7)*+{\textcolor{red}{y_1}} },
      { (-2.1, 2.77)*+{\textcolor{red}{x_2}} },
      { (-3.8, 0.7)*+{\textcolor{red}{y_2}} },
      { (-3.4, -0.7)*+{\textcolor{red}{x_3}} },
      { (-2.6, -2.77)*+{\textcolor{red}{y_3}} },
      { (2.7, 2.77)*+{\textcolor{red}{y_n}} },
      { (3.3, 0.7)*+{\textcolor{red}{x_n}} },
      { (4, 4)*+{\text{\ding{172}}} },
      { (-2, 5)*+{\text{\ding{173}}} },
      { (-5.5, 1)*+{\text{\ding{174}}} },
      }
    \endxy
  \end{equation*}
  \caption{A segment of link $L$ which is surrounded by  
    alternating  crossings is depicted.
    We number each vertex as \ding{172}, \ding{173}, $\ldots$.}
  \label{fig:alternate_knot}
\end{figure}

We first consider a surface $D_a$, which is surrounded by  alternating
crossings as in Fig.~\ref{fig:alternate_knot}.
We assign octahedron  for each crossing following
eq.~\eqref{attach_octa}, and introduce variables as
shown there.
A contribution to the  invariant $\tau_1(L)$
from above segment of link
$L$
is thus
given by
\begin{equation}
  \iiint \mathrm{d} p_1 \cdots \mathrm{d} p_n \
  \prod_{i=0}^{n-1}
  \langle p_{i+1} , x_{i+1} \  | \  S^{-1} \ | \  y_i, p_i \rangle
  ,
\end{equation}
where we use $y_0=y_n$ and $p_0=p_n$.
By substituting an asymptotic form~\eqref{S_asymptotics},
we get
$p_{i+1}= p_1 + \sum_{j=1}^{i} y_j$ for $i>0$, and the integral reduces to
\begin{equation*}
  \delta(y_1 + \dots + y_n) \cdot
  \int \mathrm{d} p_1 \
  \exp \frac{1}{2 \, \mathrm{i} \, \gamma}
  \biggl(
  \sum_{i=0}^{n-1}
  \Bigl( \frac{\pi^2}{6} - \Li(\mathrm{e}^{x_{i+1} - p_i})
  - x_{i+1} \, y_i \Bigr)
  +\sum_{1 \leq i< j \leq n} y_i \, y_j
  \biggr) .
\end{equation*}
We evaluate this integral at the saddle point, whose condition is
\begin{equation}
  \prod_{i=0}^{n-1}
  \bigl(
  1 - \mathrm{e}^{x_{i+1} - p_i}
  \bigr)
  = 1  .
\end{equation}
This equation coincides with
the hyperbolicity condition for gluing $n$
tetrahedra in surface $D_a$ along an axis parallel to  axes
$\otimes$ in Fig.~\ref{fig:alternate_knot}
(see also Fig.~\ref{fig:41_glue_1} and Fig.~\ref{fig:41_glue_sp}
for  $n=3$ and $n=2$ cases).

We can see that the same
correspondence occurs for non-alternating case.
We suppose a surface $D_a$ is surrounded like
Fig.~\ref{fig:alternate_knot} whereas  each  vertex  $i$ is
either
over-crossing $\xybox{0;/r1.4pc/:, {\xoverv<|||>{~D_a}} }$ or
under-crossing  $\xybox{0;/r1.4pc/:, {\xunderv<|||>{D_a}} }$,
which  we  denote $i \in \mathcal{O}$ and $i \in \mathcal{U}$
respectively.
In this case  a  contribution to the invariant is given by
\begin{equation}
  \iiint \mathrm{d} p_1 \cdots \mathrm{d}p_n \
  \prod_{i \in \mathcal{O}}
  \langle p_{i+1}, x_{i+1} \ | \  S^{-1} \  | \ y_i , p_i \rangle \cdot
  \prod_{i \in \mathcal{U}}
  \langle x_{i+1} , p_{i+1} \  | \  S \  | \  p_i , y_i \rangle .
\end{equation}
By substituting an expression~\eqref{S_asymptotics} 
we obtain
\begin{align*}
  p_{k+1}
  & =
  p_1 + \sum_{
    \substack{1\leq i \leq k\\
      i\in \mathcal{O}}} y_i
  - \sum_{
    \substack{1 \leq i \leq k\\
      i \in \mathcal{U}}}
  x_{i+1} ,
  &
  & \sum_{i \in \mathcal{O}} y_i = \sum_{i \in \mathcal{U}} x_{i+1} ,
\end{align*}
and the integral becomes
\begin{multline*}
  \delta(\sum_{i \in \mathcal{O}} y_i - \sum_{i \in \mathcal{U}} x_{i+1} )
  \int \mathrm{d} p_1 \
  \exp \frac{1}{2 \, \mathrm{i} \, \gamma}
  \biggl(
  \sum_{i\in \mathcal{O}}
  \Bigl(
  \frac{\pi^2}{6} - \Li(\mathrm{e}^{x_{i+1} - p_i}) - x_{i+1} \, y_i
  \Bigr)
  \\
  +
  \sum_{i \in \mathcal{U}}
  \Bigl(
  -\frac{\pi^2}{6} + \Li(\mathrm{e}^{y_i - p_{i+1}}) + x_{i+1} \, y_i
  \Bigr)
  +
  \sum_{\substack{i,j \in \mathcal{O}\\
    i<j}} y_i \, y_j
  -
  \sum_{\substack{i \in \mathcal{U} \\
      j \in \mathcal{O}}} x_{i+1} \, y_j
  +
  \sum_{\substack{i,j \in \mathcal{U}\\
      i \leq j}}
  x_{i+1} \, x_{j+1}
  \biggr).
\end{multline*}
The saddle point equation is given as
\begin{equation}
  \prod_{i \in \mathcal{O}}
  \Bigl(
  1- \mathrm{e}^{x_{i+1} - p_i}
  \Bigr)
  \cdot
  \prod_{i \in \mathcal{U}}
  \Bigl(
  1- \mathrm{e}^{y_i - p_{i+1}}
  \Bigr)^{-1}
  =1 ,
\end{equation}
which coincides with  the hyperbolicity equation around an axis in
$D_a$   parallel to axes $\otimes$.

\subsection{Gluing Around Ridgeline of Octahedron}

\begin{figure}[htbp]
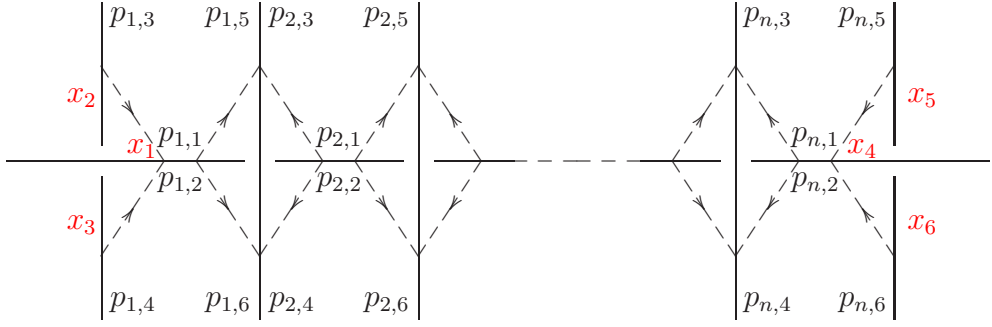

  \begin{equation*}
    \xy/r1pc/:.
    {
      {  (2,0) \ar@{-}|@{} (9.5,0) },
      {  (10.5,0) \ar@{-}|@{} (14.5,0) },
      {  (15.5,0) \ar@{-}|@{} (18,0) },
      {  (18,0) \ar@{--}|@{} (22,0) },
      {  (22,0) \ar@{-}|@{} (24.5,0) },
      {  (25.5,0) \ar@{-}|@{} (33,0) },
      {  (5,5) \ar@{-}|@{} (5,0.5) },
      {  (5,-0.5) \ar@{-}|@{} (5,-5) },
      {  (30,5) \ar@{-}|@{} (30,0.5) },
      {  (30,-0.5) \ar@{-}|@{} (30,-5) },
      {  (10,5) \ar@{-}|@{} (10,-5) },
      {  (15,5) \ar@{-}|@{} (15,-5) },
      {  (25,5) \ar@{-}|@{} (25,-5) },
      { (7.5,0.7)*+{p_{1,1}} },
      { (7.5,-0.7)*+{p_{1,2}} },
      { (6,4.5)*+{p_{1,3}} },
      { (6,-4.5)*+{p_{1,4}} },
      { (9,4.5)*+{p_{1,5}} },
      { (9,-4.5)*+{p_{1,6}} },
      { (12.5,0.7)*+{p_{2,1}} },
      { (12.5,-0.7)*+{p_{2,2}} },
      { (11,4.5)*+{p_{2,3}} },
      { (11,-4.5)*+{p_{2,4}} },
      { (14,4.5)*+{p_{2,5}} },
      { (14,-4.5)*+{p_{2,6}} },
      { (27.5,0.7)*+{p_{n,1}} },
      { (27.5,-0.7)*+{p_{n,2}} },
      { (26,4.5)*+{p_{n,3}} },
      { (26,-4.5)*+{p_{n,4}} },
      { (29,4.5)*+{p_{n,5}} },
      { (29,-4.5)*+{p_{n,6}} },
      { (7, 0) \ar@{--}|@{<} (5,3)  },
      { (7, 0) \ar@{--}|@{<} (5,-3)  },
      { (8, 0) \ar@{--}|@{>} (10,3)  },
      { (8, 0) \ar@{--}|@{>} (10,-3)  },
      { (12, 0) \ar@{--}|@{>} (10,3)  },
      { (12, 0) \ar@{--}|@{>} (10,-3)  },
      { (13, 0) \ar@{--}|@{>} (15,3)  },
      { (13, 0) \ar@{--}|@{>} (15,-3)  },
      { (17, 0) \ar@{--}|@{>} (15,3)  },
      { (17, 0) \ar@{--}|@{>} (15,-3)  },
      { (23, 0) \ar@{--}|@{>} (25,3)  },
      { (23, 0) \ar@{--}|@{>} (25,-3)  },
      { (27, 0) \ar@{--}|@{>} (25,3)  },
      { (27, 0) \ar@{--}|@{>} (25,-3)  },
      { (28, 0) \ar@{--}|@{<} (30,3)  },
      { (28, 0) \ar@{--}|@{<} (30,-3)  },
      { (6.4,0.4)*+{\text{\textcolor{red}{$x_1$} }} },
      { (4.5,2)*+{\text{\textcolor{red}{$x_2$} }} },
      { (4.5,-2)*+{\text{\textcolor{red}{$x_3$} }} },
      { (29.1,0.4)*+{\text{\textcolor{red}{$x_4$} }} },
      { (31,2)*+{\text{\textcolor{red}{$x_5$} }} },
      { (31,-2)*+{\text{\textcolor{red}{$x_6$} }} },
      }
    \endxy
  \end{equation*}
  \caption{We parameterize a segment of link $L$.}
  \label{fig:Ridge_condition}
\end{figure}

We shall check  a correspondence between the hyperbolic condition and the
saddle point equation for ridgelines of octahedron.
We consider a case such as Fig.~\ref{fig:Ridge_condition}.
Therein
$n-1$ over-crossings $\text{\rotatebox{45}{$\xybox{0;/r1.4pc/:, {\xoverv}}$}}$
are  sandwiched by two under-crossings
$\text{\rotatebox{45}{$\xybox{0;/r1.4pc/:, {\xunderv}}$}}$.

\begin{figure}[htbp]
  \begin{center}
    \parbox{5cm}{
      \begin{psfrags}
        \psfrag{A}{\rotatebox{20}{$p_{1,1}$}}
        \psfrag{B}{\rotatebox{0}{$p_{1,2}$}}
        \psfig{file=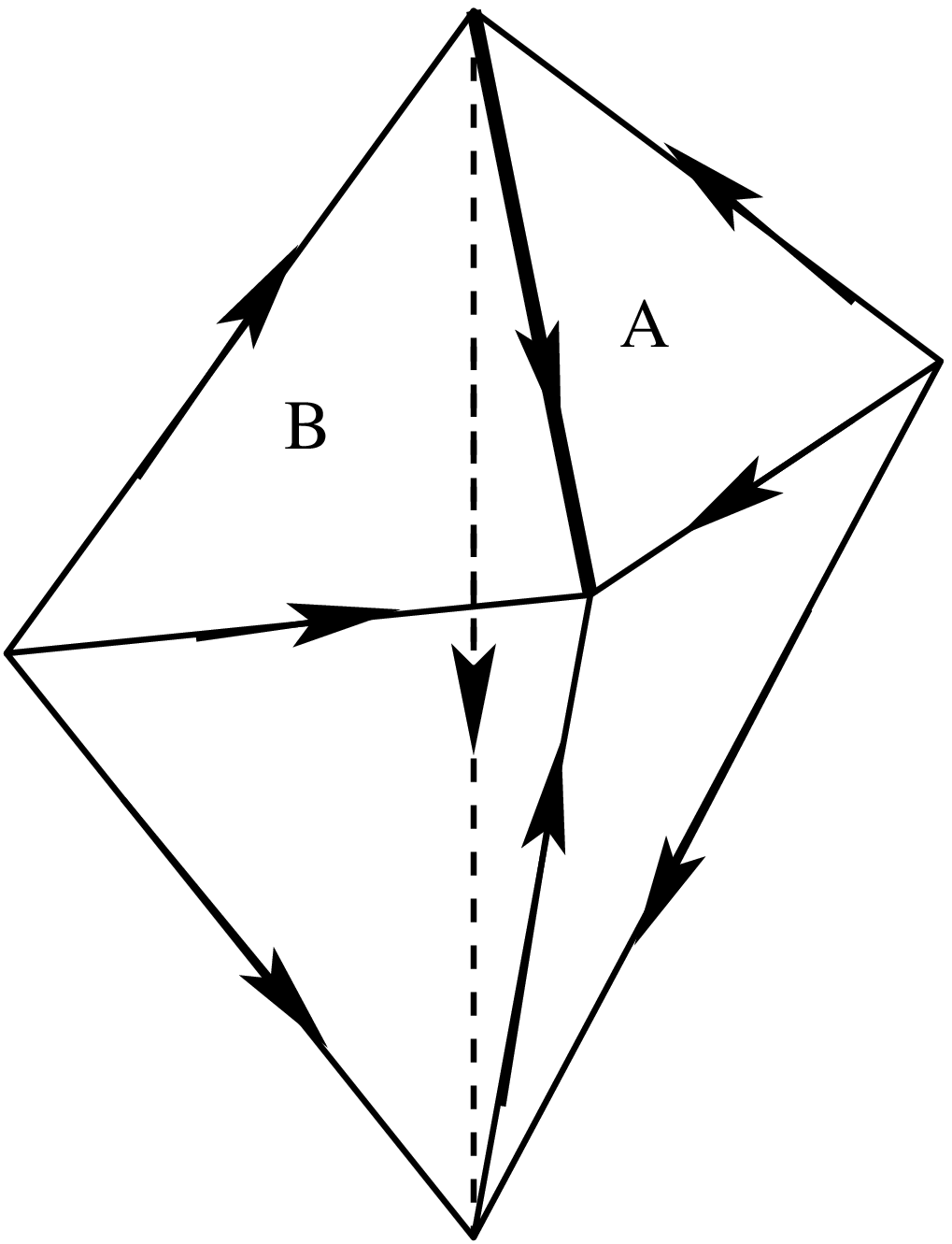,scale=0.4} 
      \end{psfrags}
      }
    \parbox{5cm}{
      \begin{psfrags}
        \psfrag{A}{\rotatebox{-10}{\rotatebox{10}{$p_{j,2}$}}}
        \psfrag{B}{\textcolor{red}{\reflectbox{\rotatebox{-20}{$p_{j,1}$}}}}
        \psfrag{C}{\rotatebox{20}{$p_{j+1,2}$}}
        \psfrag{D}{\textcolor{red}{\reflectbox{\rotatebox{30}{$p_{j+1,1}$}}}}
        \psfrag{E}{$p_{j,6}$}
        \psfrag{F}{\textcolor{red}{\reflectbox{\rotatebox{-10}{$p_{j,5}$}}}}
        \psfrag{G}{\rotatebox{20}{$p_{j+1,4}$}}
        \psfrag{H}{\textcolor{red}{\reflectbox{\rotatebox{10}{$p_{j+1,3}$}}}}
        \psfig{file=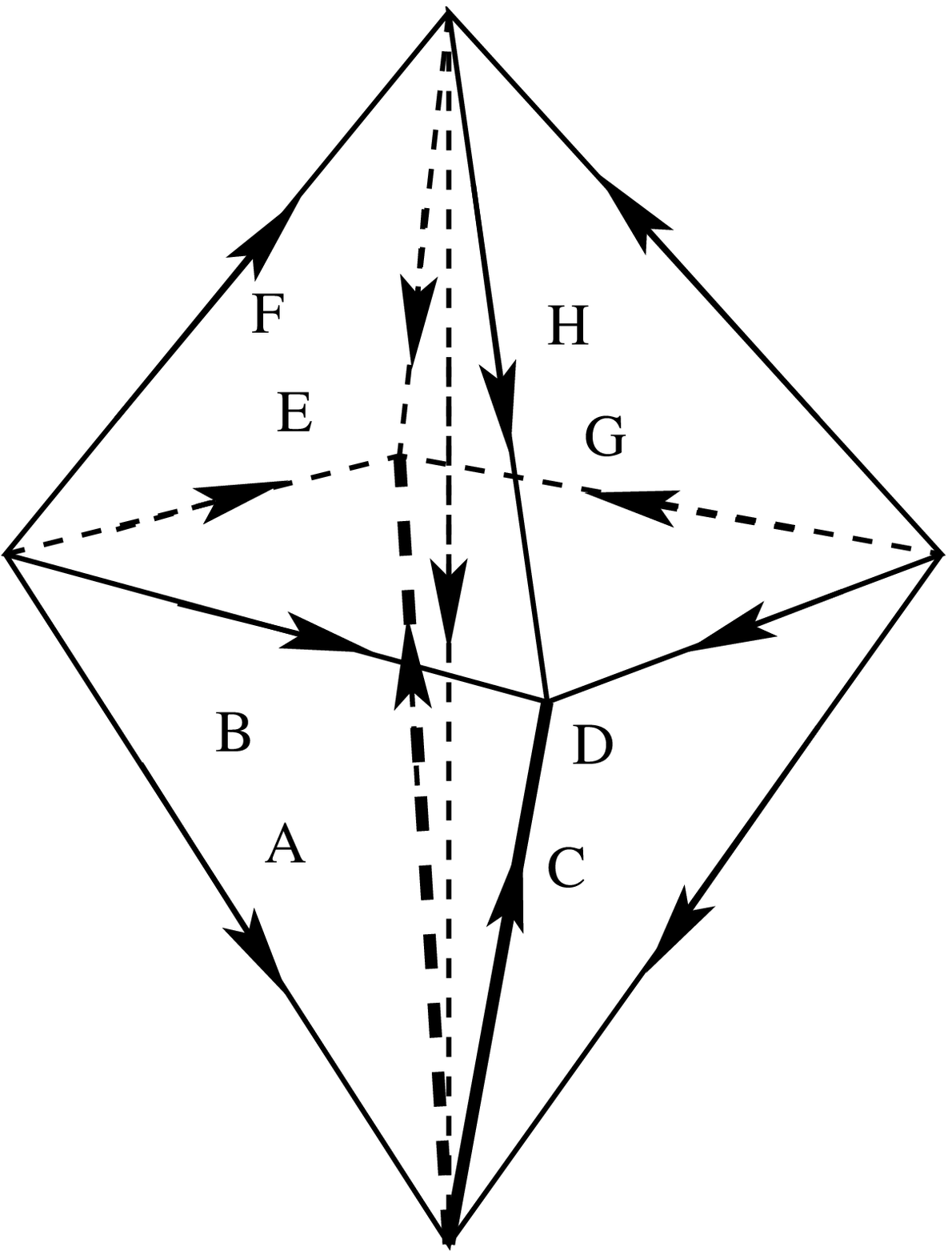,scale=0.35}
      \end{psfrags}
      }
    \parbox{4cm}{
      \begin{psfrags}
        \psfrag{A}{\rotatebox{-20}{$p_{n,1}$}}
        \psfrag{B}{\rotatebox{10}{$p_{n,2}$}}
        \psfig{file=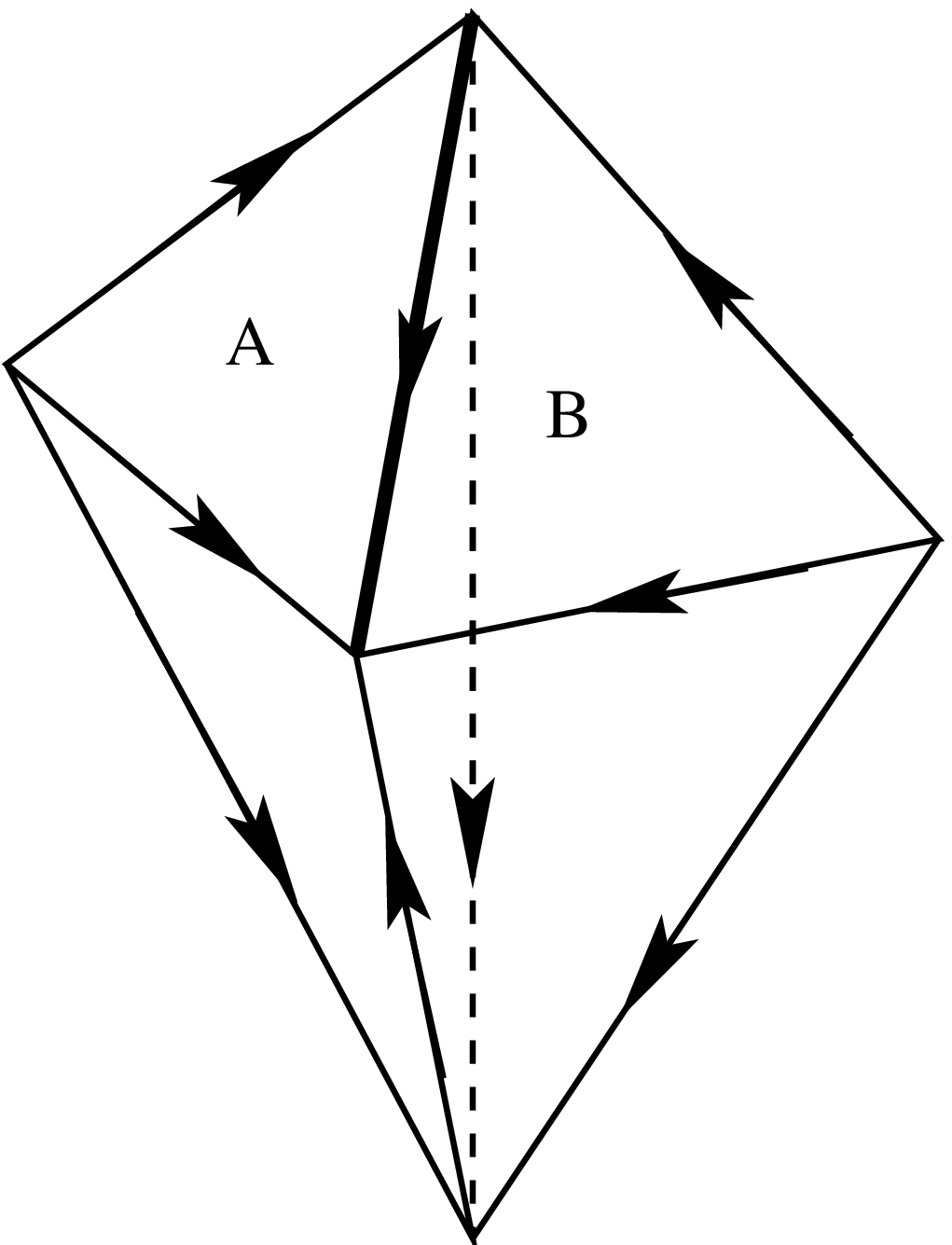,scale=0.4}
      \end{psfrags}
      }
    \caption{
      The 3-dimensional picture of
      Fig.~\ref{fig:Ridge_condition} is given.
      Eq.~\eqref{hyperbolic_ridge} coincides with the hyperbolicity
      condition around  ridgelines (bold line)
      of the octahedra.
      }
    \label{fig:Ridge_octa}
  \end{center}
\end{figure}

A contribution from this segment is given by
\begin{multline}
  \iiint \prod_{i=1}^n \mathrm{d} p_{i,1} \ \mathrm{d} p_{i,2} \cdot
  \mathrm{d} x_1 \   \mathrm{d} x_4 \
  \langle x_1 , p_{1,1} \  | \  S \  | \  p_{1,3} , x_2 \rangle \,
  \langle p_{1,4}, x_3 \ | \  S^{-1} \  | \  x_1, p_{1,2}  \rangle
  \\
  \times
  \prod_{j=1}^{n-1}
  \langle p_{j,5}, p_{j+1,3} , p_{j+1,1}, p_{j+1,2} \  | \
  \Check{R} \ | \
  p_{j,1} , p_{j,2} , p_{j,6} , p_{j+1,4} \rangle \\
  \times
  \langle p_{n,5} , x_5 \  | \  S^{-1} \  | \ x_4 , p_{n,1} \rangle \,
  \langle x_4 , p_{n,2} \ | \ S \  | \  p_{n,6} , x_6 \rangle .
\end{multline}
We substitute eqs.~\eqref{S_asymptotics} and~\eqref{R_asymptotics} into
above equation.
We get
\begin{align*}
  p_{j,1} &= p_{j+1,1} - p_{j+1,3} + p_{j,5},
  &
  p_{j,2} &= p_{j+1,2} - p_{j+1,4} + p_{j,6},
  \\[2mm]
  p_{n,1} & = - x_4 + p_{n,5} ,
  &
  p_{n,2} &= - x_4 + p_{n,6} ,
  \\
  x_1 &= x_4 + \sum_{j=1}^n ( p_{j,3} - p_{j,5} ) ,
\end{align*}
and the integral reduces to that of $x_4$-integration,
whose saddle point equation is given by
\begin{equation*}
  \frac{
    (1 - \mathrm{e}^{x_3 - p_{1,2}} ) \,
    ( 1 - \mathrm{e}^{x_5 - p_{n,1}} )
    }{
    (1 - \mathrm{e}^{x_2 - p_{1,1}} ) \,
    ( 1 - \mathrm{e}^{x_6 - p_{n,2}} )
    }
  \cdot
  \mathrm{e}^{
    + x_2 - x_3 - x_5 + x_6}
  =1 .
\end{equation*}
One sees that this equation coincides with the hyperbolicity equation
around a ridgeline of the octahedron
(bold lines in Fig~\ref{fig:Ridge_octa});
\begin{equation}
  \label{hyperbolic_ridge}
  \frac{1-\mathrm{e}^{p_{1,1} - x_2}}{1-\mathrm{e}^{p_{1,2} - x_3}}
  \cdot
  \prod_{j=1}^{n-1}
  \mathrm{e}^{-p_{j,5} + p_{j,6} + p_{j+1,3} - p_{j+1,4}}
  \cdot
  \frac{
    1-\mathrm{e}^{p_{n,2} - x_6}}{
    1 - \mathrm{e}^{p_{n,1} - x_5}
    } 
  = 1.
\end{equation}

\begin{figure}[htbp]
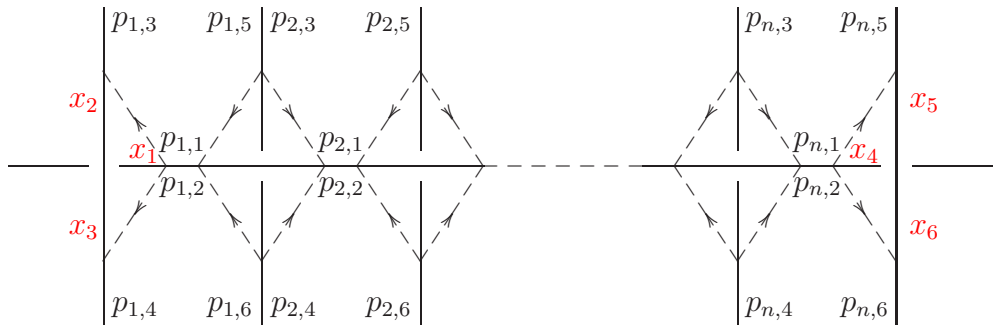

  \begin{equation*}
    \xy/r1pc/:.
    {
      { (-3,0)\ar@{-} (-0.5,0)},
      { (0.5,0)\ar@{-} (12,0)},
      { (12,0)\ar@{--} (17,0)},
      { (17,0)\ar@{-} (24.5,0)},
      { (25.5,0)\ar@{-} (28,0)},
      { (0,5)\ar@{-} (0,-5)},
      { (25,5)\ar@{-} (25,-5)},
      { (5,5)\ar@{-} (5,0.5)},
      { (5,-0.5)\ar@{-} (5,-5)},
      { (10,5)\ar@{-} (10,0.5)},
      { (10,-0.5)\ar@{-} (10,-5)},
      { (20,5)\ar@{-} (20,0.5)},
      { (20,-0.5)\ar@{-} (20,-5)},
      { (2.5,0.7)*+{p_{1,1}} },
      { (2.5,-0.7)*+{p_{1,2}} },
      { (1,4.5)*+{p_{1,3}} },
      { (1,-4.5)*+{p_{1,4}} },
      { (4,4.5)*+{p_{1,5}} },
      { (4,-4.5)*+{p_{1,6}} },
      { (7.5,0.7)*+{p_{2,1}} },
      { (7.5,-0.7)*+{p_{2,2}} },
      { (6,4.5)*+{p_{2,3}} },
      { (6,-4.5)*+{p_{2,4}} },
      { (9,4.5)*+{p_{2,5}} },
      { (9,-4.5)*+{p_{2,6}} },
      { (22.5,0.7)*+{p_{n,1}} },
      { (22.5,-0.7)*+{p_{n,2}} },
      { (21,4.5)*+{p_{n,3}} },
      { (21,-4.5)*+{p_{n,4}} },
      { (24,4.5)*+{p_{n,5}} },
      { (24,-4.5)*+{p_{n,6}} },
      { (2, 0) \ar@{--}|@{>} (0,3)  },
      { (2, 0) \ar@{--}|@{>} (0,-3)  },
      { (3, 0) \ar@{--}|@{<} (5,3)  },
      { (3, 0) \ar@{--}|@{<} (5,-3)  },
      { (7, 0) \ar@{--}|@{<} (5,3)  },
      { (7, 0) \ar@{--}|@{<} (5,-3)  },
      { (8, 0) \ar@{--}|@{<} (10,3)  },
      { (8, 0) \ar@{--}|@{<} (10,-3)  },
      { (12, 0) \ar@{--}|@{<} (10,3)  },
      { (12, 0) \ar@{--}|@{<} (10,-3)  },
      { (18, 0) \ar@{--}|@{<} (20,3)  },
      { (18, 0) \ar@{--}|@{<} (20,-3)  },
      { (22, 0) \ar@{--}|@{<} (20,3)  },
      { (22, 0) \ar@{--}|@{<} (20,-3)  },
      { (23, 0) \ar@{--}|@{>} (25,3)  },
      { (23, 0) \ar@{--}|@{>} (25,-3)  },
      { (1.4,0.4)*+{\text{\textcolor{red}{$x_1$} }} },
      { (-0.5,2)*+{\text{\textcolor{red}{$x_2$} }} },
      { (-0.5,-2)*+{\text{\textcolor{red}{$x_3$} }} },
      { (24.1,0.4)*+{\text{\textcolor{red}{$x_4$} }} },
      { (26,2)*+{\text{\textcolor{red}{$x_5$} }} },
      { (26,-2)*+{\text{\textcolor{red}{$x_6$} }} },
      }
    \endxy
  \end{equation*}
  \caption{A segment of link $L$.}
  \label{fig:Ridge_2}
\end{figure}

In the same manner, we can see a correspondence between the    saddle
point equation and the hyperbolicity condition in a case
that
$n-1$ under-crossings $\text{\rotatebox{45}{$\xybox{0;/r1.4pc/:, {\xunderv}}$}}$
are  sandwiched by two over-crossings
$\text{\rotatebox{45}{$\xybox{0;/r1.4pc/:, {\xoverv}}$}}$.
(Fig.~\ref{fig:Ridge_2}).

\begin{figure}[htbp]
  \begin{center}
    \parbox{5cm}{
      \begin{psfrags}
        \psfrag{A}{\rotatebox{20}{$p_{1,1}$}}
        \psfrag{B}{\rotatebox{0}{$p_{1,2}$}}
        \psfig{file=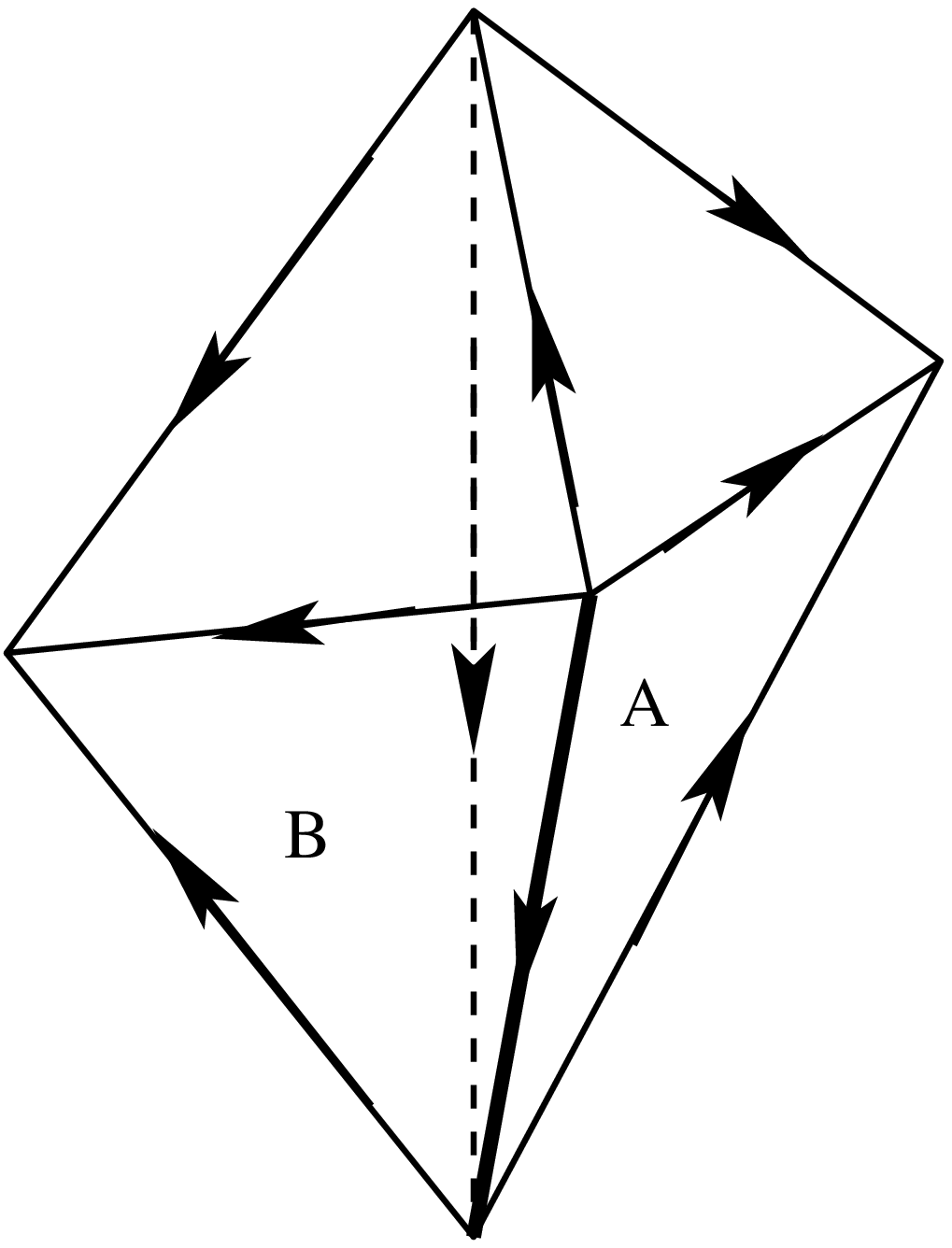,scale=0.4} 
      \end{psfrags}
      }
    \parbox{5cm}{
      \begin{psfrags}
        \psfrag{A}{\rotatebox{-10}{$p_{j,6}$}}
        \psfrag{B}{\textcolor{red}{\reflectbox{\rotatebox{-20}{$p_{j,5}$}}}}
        \psfrag{C}{\rotatebox{20}{$p_{j+1,4}$}}
        \psfrag{D}{\textcolor{red}{\reflectbox{\rotatebox{30}{$p_{j+1,3}$}}}}
        \psfrag{E}{$p_{j,2}$}
        \psfrag{F}{\textcolor{red}{\reflectbox{\rotatebox{-10}{$p_{j,1}$}}}}
        \psfrag{G}{\rotatebox{20}{$p_{j+1,2}$}}
        \psfrag{H}{\textcolor{red}{\reflectbox{\rotatebox{10}{$p_{j+1,1}$}}}}
        \psfig{file=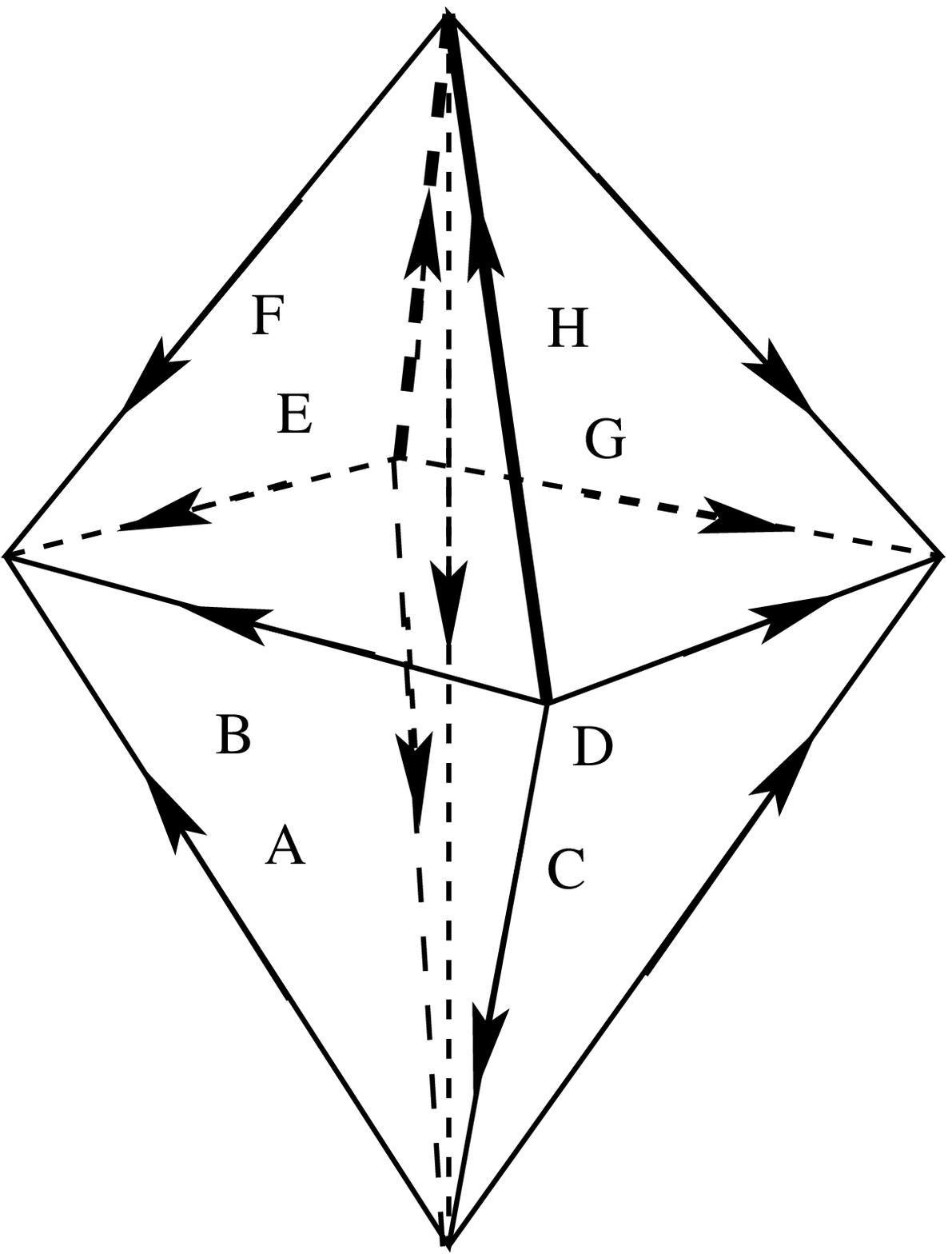,scale=0.35}
      \end{psfrags}
      }
    \parbox{4cm}{
      \begin{psfrags}
        \psfrag{A}{\rotatebox{-20}{$p_{n,1}$}}
        \psfrag{B}{\rotatebox{10}{$p_{n,2}$}}
        \psfig{file=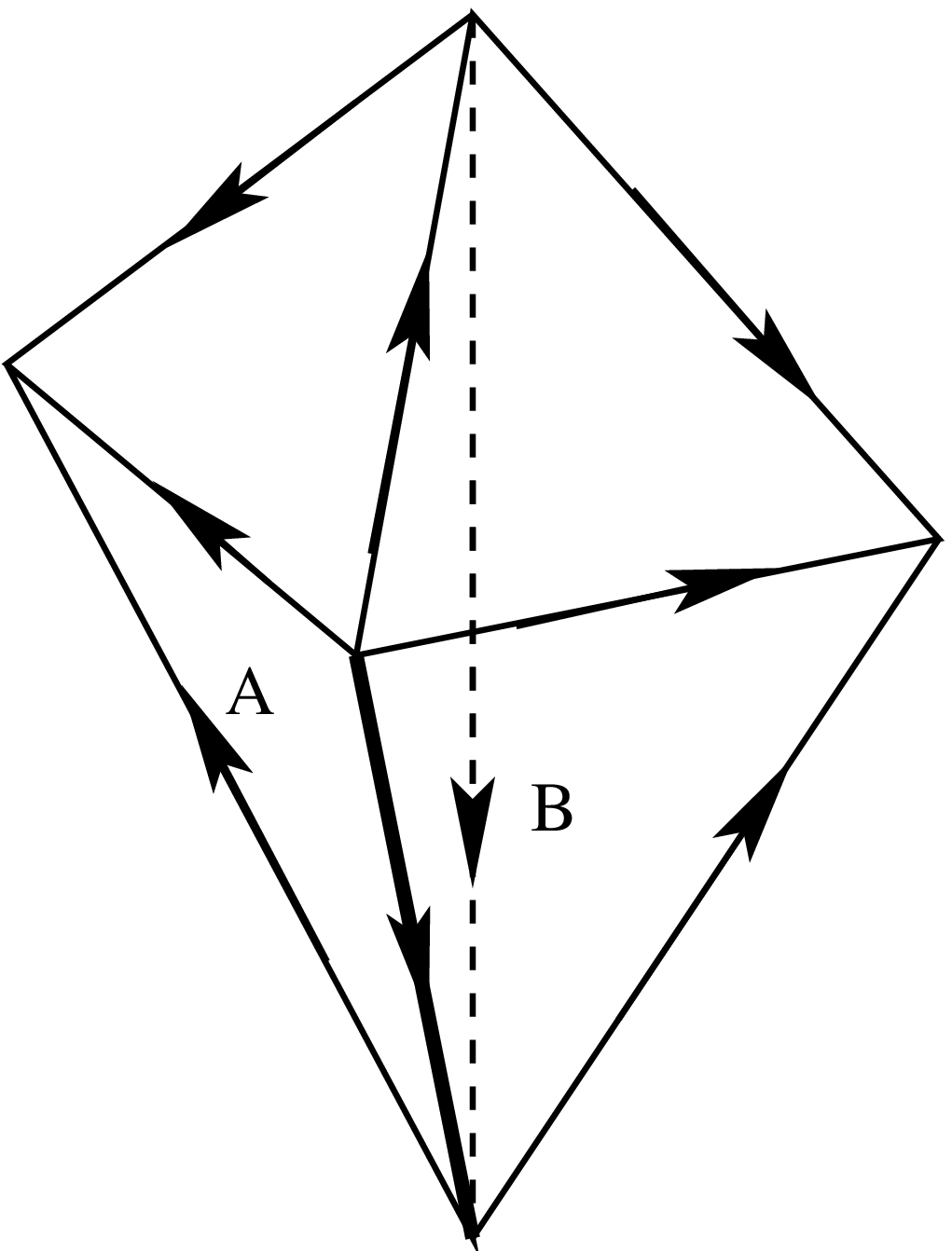,scale=0.4}
      \end{psfrags}
      }
  \end{center}
  \caption{
    Depicted are  polyhedra
    which  come  from Fig.~\ref{fig:Ridge_2}.
    The hyperbolicity condition around bold edges is given in
    eq.~\eqref{condition_ridge_2_octa}.}
  \label{fig:Ridge_2_octa}
\end{figure}

In this case a contribution to the invariant is given by the integral,
\begin{multline}
  \iiint \prod_{j=1}^n \mathrm{d} p_{j,1} \ \mathrm{d} p_{j,2} \cdot
  \mathrm{d} x_1 \ \mathrm{d} x_4 \
  \langle p_{1,1} , x_1 \  | \  S^{-1} \  | \ x_2 , p_{1,3} \rangle \,
  \langle x_3 , p_{1,4} \ | \  S \  | \ p_{1,2} , x_1 \rangle
  \\
  \times
  \prod_{j=1}^{n-1}
  \langle p_{j,2} , p_{j,1} , p_{j,5}, p_{j+1,3} \  | \
  \Check{R}  \ | \
  p_{j,6} , p_{j+1,4}, p_{j+1,2} , p_{j+1,1} \rangle
  \\
  \times
  \langle x_5 , p_{n,5} \  | \  S \ | \  p_{n,1} , x_4 \rangle \,
  \langle p_{n,2} , x_4 \ | \ S^{-1} \  | \  x_6 , p_{n,6} \rangle .
\end{multline}
By substituting eqs.~\eqref{S_asymptotics} and~\eqref{R_asymptotics}, 
we obtain
\begin{align*}
  p_{j,2} & = p_{j,1} - p_{j,5} + p_{j,6} ,
  &
  p_{1,1} & = p_{1,3} + x_2 ,
  \\[2mm]
  p_{n,1} & = x_5 + p_{n,5},
  &
  p_{n,2} &= x_6 + p_{n,6} ,
\end{align*}
and the integral reduces to an integration over
$x_1$, $x_4$, and $p_{j,1}$ for
$j=2, 3, \dots, n-1$.
The saddle point equations for $x_1$ and $x_4$ are respectively
written as
\begin{align*}
  \frac{1 - \mathrm{e}^{x_1 - p_{1,3}}}
  {1 - \mathrm{e}^{x_1 -  p_{1,4}}}
  \cdot
  \mathrm{e}^{-x_2 + x_3}
  & = 1 ,
  &
  \frac{1 - \mathrm{e}^{x_4 - p_{n,6}}}
  {1 - \mathrm{e}^{x_4 - p_{n,5}}}
  \cdot
  \mathrm{e}^{- x_6 + x_5}
  &  = 1 .
\end{align*}
These two equations give
\begin{equation}
  \label{condition_ridge_2_octa}
  \frac{
    1 - \mathrm{e}^{p_{1,4} - x_1}
    }{
    1 - \mathrm{e}^{p_{1,3} - x_1}
    } \cdot
  \prod_{j=1}^{n-1}
  \mathrm{e}^{-p_{j,2} + p_{j+1,2} - p_{j+1,1} + p_{j,1}}
  \cdot
  \frac{
    1 - \mathrm{e}^{p_{n,5} - x_4}
    }{
    1 - \mathrm{e}^{p_{n,6} - x_4}
    }
  = 1 ,
\end{equation}
which denotes the hyperbolicity equation around  ridgelines of the
octahedron
(Fig.~\ref{fig:Ridge_2_octa}).

As a result,
we have seen that
the 3-dimensional hyperbolic structure naturally appears in
the invariant $\Tilde{\tau}_1(K)$, \emph{i.e.},
a classical limit of
the knot invariant $\tau_1(K)$  which is defined
by the integral form based on the quantum dilogarithm function.
The saddle point equation  exactly coincides
with the hyperbolicity consistency  condition in  gluing the octahedra which is
assigned to each crossing.
To be precise,
in order to see that
a finite collection of ideal tetrahedra results in  a 3-manifold,
we need to  prove  the completeness condition
by showing that
the  developing map near the ideal vertex yields Euclidean structure.
We have  checked this condition for several knots,
but  we do not
have proof at this moment.
There is still another problem to be solved.
Generally a set of the saddle point equations (hyperbolicity
consistency conditions)  has several algebraic solutions, and
we are not sure which solutions
among   them
we should choose  as 
dominant  in a definition of the invariant $\Tilde{\tau}_1(K)$.
When we assume that  
geometrically preferable   solutions $z_i$ of the saddle point
conditions
are
dominant
in  the  classical limit,
we may  conclude that
\begin{align}
  \label{result_tau}
  \Tilde{\tau}_1(K)
  & =
  \sum_{\text{ideal tetrahedra}:  i}
  L(1-z_i) ,
%
\end{align}
as each tetrahedron has a function $V(x,y)$
which reduces to the Rogers dilogarithm
function   at the critical point~\eqref{property_V}.


\section{Example: Figure-Eight Knot}
\label{sec:eight}

\begin{figure}[htbp]
  \begin{center}
    \psfig{file=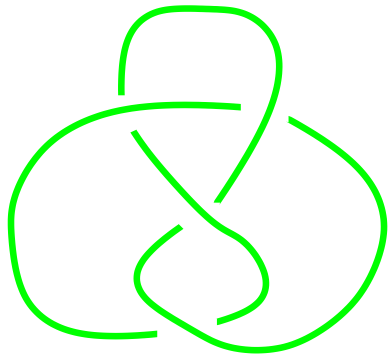,scale=0.25}
    \caption{The figure-eight knot $4_1$.}
    \label{fig:fig_eight}
  \end{center}
\end{figure}

\begin{figure}[htbp]
  \begin{center}
    \begin{psfrags}
      \psfrag{A}{$D_0$}
      \psfrag{B}{$D_1$}
      \psfrag{C}{$D_2$}
      \psfrag{D}{$D_3$}
      \psfrag{E}{$D_4$}
      \psfrag{F}{$D_5$}
      \psfrag{1}{\ding{172}}
      \psfrag{2}{\ding{173}}
      \psfrag{3}{\ding{174}}
      \psfrag{4}{\ding{175}}
      \psfig{file=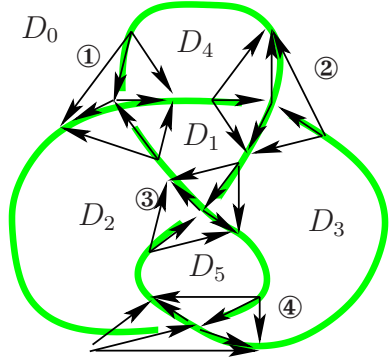,scale=0.25}      
    \end{psfrags}
    \caption{We attach octahedra for each crossing of the figure-eight
      knot.}
    \label{fig:fig_eight_tetra}
  \end{center}
\end{figure}

We shall demonstrate
how to decompose the knot complement  into  tetrahedra in a
case of the  figure-eight knot.
The figure-eight knot is
given as $\sigma_1 \, \sigma_2^{~-1} \, \sigma_1
\, \sigma_2^{~-1}$ in the braid group, and is
depicted as Fig.~\ref{fig:fig_eight}.
To each crossing in the  figure-eight knot,  we assign the octahedron
(Fig.~\ref{fig:R_matrix}
or eq.~\eqref{attach_octa}),
and
give the numbering to each crossing  as in
Fig.~\ref{fig:fig_eight_tetra}.
We have  also named  each surface as  $D_i$.
We call the surface inside the octahedron as $S^a_{i,j}$;
a  surface  is  in  the octahedron of the $a$-th crossing,
and is a boundary
between $D_i$ and $D_j$.

\begin{figure}[htbp]
  \begin{center}
    \parbox{8cm}{
      \begin{psfrags}
        \psfrag{A}{$S^1_{1,2}$}
        \psfrag{B}{\textcolor{red}{\reflectbox{$S^1_{1,4}$}}}
        \psfrag{C}{$S^3_{1,3}$}
        \psfrag{D}{$S^3_{1,2}$}
        \psfrag{E}{\textcolor{red}{\reflectbox{$S^2_{1,4}$}}}
        \psfrag{F}{$S^2_{1,3}$}
        \psfrag{X}{\ding{172}}
        \psfrag{Y}{\ding{173}}
        \psfrag{Z}{\ding{174}}
        \psfig{file=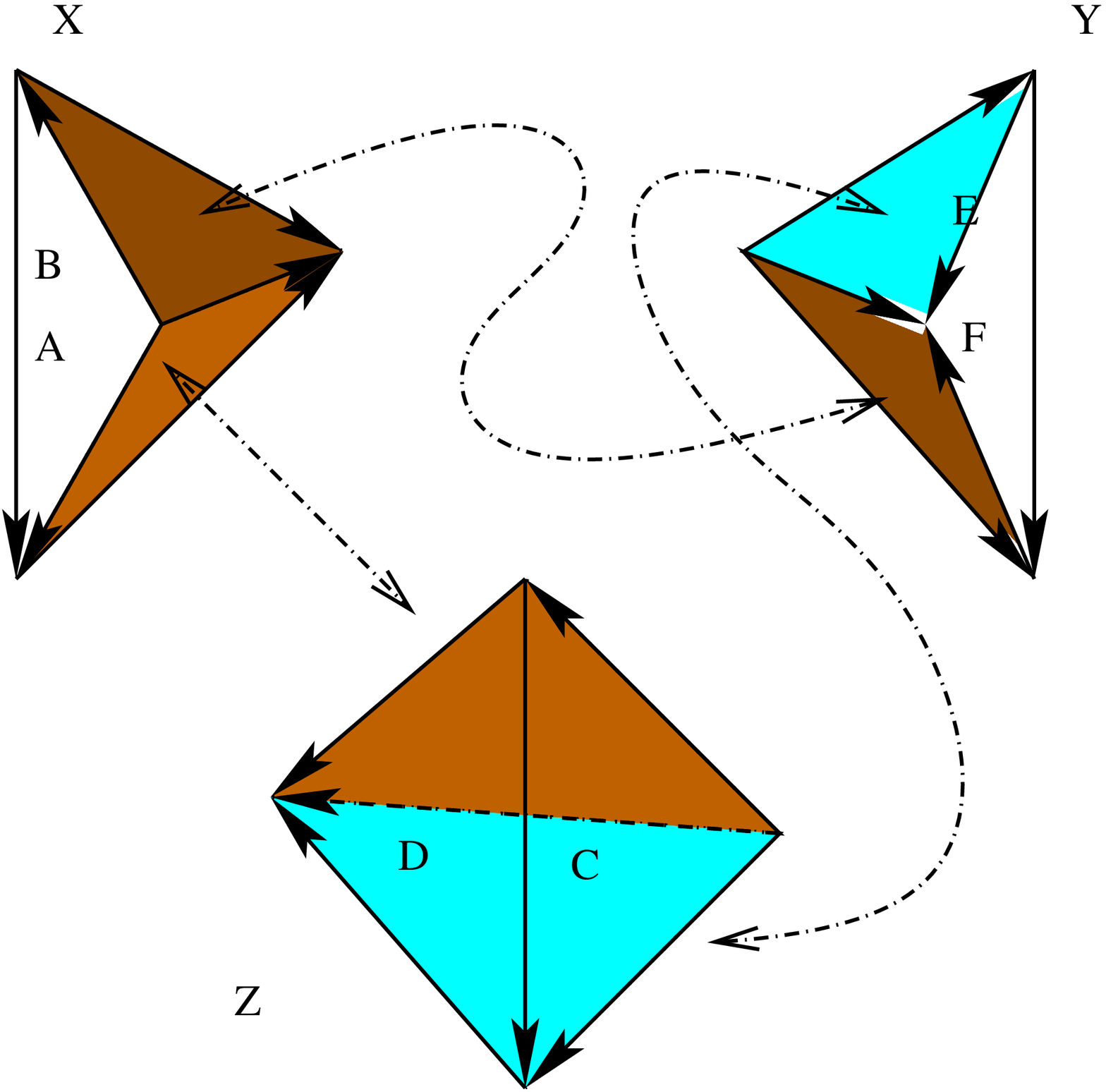,scale=0.32}
      \end{psfrags}
      }
    ~~~$\Longleftrightarrow$~~~
    \parbox{5cm}{
      \begin{psfrags}
        \psfrag{A}{\rotatebox{-30}{$S^1_{1,4}$}}
        \psfrag{B}{\rotatebox{30}{$S^2_{1,3}$}}
        \psfrag{C}{\textcolor{red}{\reflectbox{$S^3_{1,2}$}}}
        \psfrag{D}{\rotatebox{-30}{$S^1_{1,2}$}}
        \psfrag{E}{\rotatebox{30}{$S^2_{1,4}$}}
        \psfrag{F}{\textcolor{red}{\reflectbox{$S^3_{1,3}$}}}
        \psfig{file=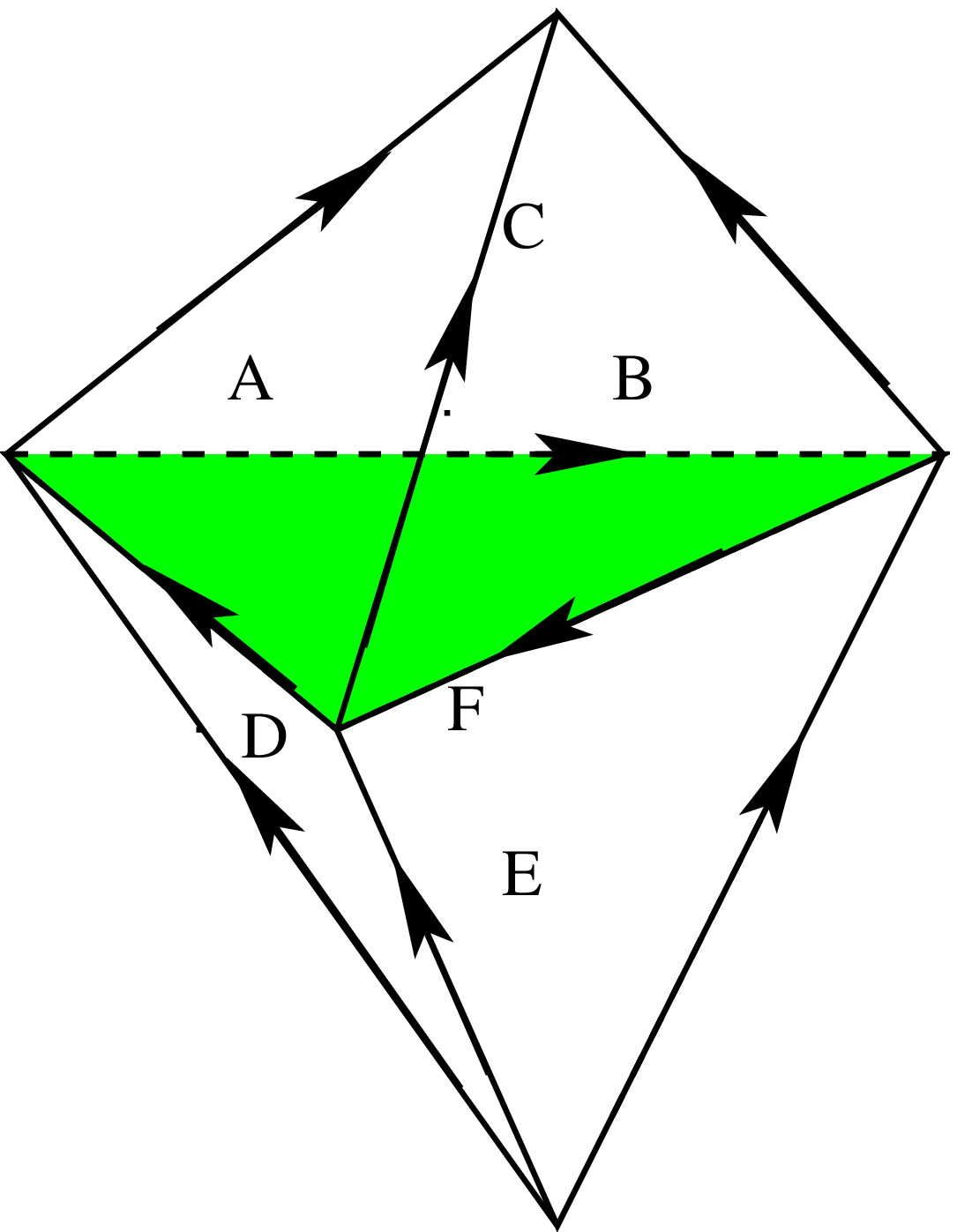,scale=0.36}
      \end{psfrags}
      }
    \caption{Gluing  3 tetrahedra in surface $D_1$ results in 2
      tetrahedra by the 2-3 Pachner move.
      The gray surface in the 2 adjacent tetrahedra is called $P_1$.
      }
    \label{fig:41_glue_1}
  \end{center}
\end{figure}

In a surface $D_1$ there are 3 tetrahedra.
These three tetrahedra are glued to each other as shown in
Fig.~\ref{fig:41_glue_1}.
Due to the pentagon
relation we obtain  2 adjacent tetrahedra, whose common surface
(gray surface)  is named  $P_1$.

\begin{figure}[htbp]
  \begin{center}
    (a)
    \parbox{4.5cm}{
      \begin{psfrags}
        \psfrag{A}{\rotatebox{-30}{$S^4_{0,3}$}}
        \psfrag{B}{\rotatebox{30}{$S^2_{0,4}$}}
        \psfrag{C}{\textcolor{red}{\reflectbox{$S^1_{0,2}$}}}
        \psfrag{D}{\rotatebox{-30}{$S^4_{0,2}$}}
        \psfrag{E}{\rotatebox{30}{$S^2_{0,3}$}}
        \psfrag{F}{\textcolor{red}{\reflectbox{$S^1_{0,4}$}}}
        \psfig{file=glue_tetra_reg.eps,scale=0.36}
      \end{psfrags}
      }
    (b)~\parbox{4.5cm}{
      \begin{psfrags}
        \psfrag{A}{\rotatebox{-30}{$S^3_{1,2}$}}
        \psfrag{B}{\rotatebox{30}{$S^4_{2,5}$}}
        \psfrag{C}{\textcolor{red}{\reflectbox{$S^1_{0,2}$}}}
        \psfrag{D}{\rotatebox{-30}{$S^3_{2,5}$}}
        \psfrag{E}{\rotatebox{30}{$S^4_{0,2}$}}
        \psfrag{F}{\textcolor{red}{\reflectbox{$S^1_{1,2}$}}}
        \psfig{file=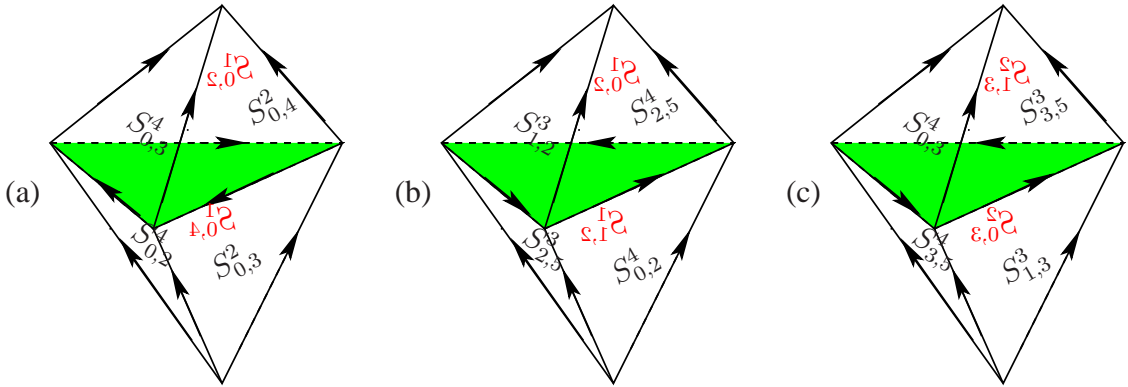,scale=0.36}
      \end{psfrags}
      }
    (c)~\parbox{4.5cm}{
      \begin{psfrags}
        \psfrag{A}{\rotatebox{-30}{$S^4_{0,3}$}}
        \psfrag{B}{\rotatebox{30}{$S^3_{3,5}$}}
        \psfrag{C}{\textcolor{red}{\reflectbox{$S^2_{1,3}$}}}
        \psfrag{D}{\rotatebox{-30}{$S^4_{3,5}$}}
        \psfrag{E}{\rotatebox{30}{$S^3_{1,3}$}}
        \psfrag{F}{\textcolor{red}{\reflectbox{$S^2_{0,3}$}}}
        \psfig{file=glue_tetra_res.eps,scale=0.36}
      \end{psfrags}
      }
    \caption{
      Gluing 3 tetrahedra in surfaces (a) $D_0$, (b) $D_2$,  and (c)
      $D_3$ gives the 2 adjacent tetrahedra.
      We call each gray
      surface  respectively $P_0$, $P_2$, and
      $P_3$.
      }
    \label{fig:41_glue_2}
  \end{center}
\end{figure}

Three  tetrahedra in a  surface $D_0$ also gives same adjacent  tetrahedra
with that in
Fig.~\ref{fig:41_glue_1}.
By the same method of gluing 3 tetrahedra in surfaces $D_2$ and $D_3$,
we obtain adjacent  tetrahedra as shown in Fig.~\ref{fig:41_glue_2}.

\begin{figure}[htbp]
  \begin{center}
    \parbox{8.5cm}{
      \begin{psfrags}
        \psfrag{A}{$S^1_{1,4}$}
        \psfrag{B}{\textcolor{red}{\reflectbox{$S^1_{0,4}$}}}
        \psfrag{C}{$S^2_{1,4}$}
        \psfrag{D}{\textcolor{red}{\reflectbox{$S^2_{0,4}$}}}
        \psfrag{X}{\ding{172}}
        \psfrag{Y}{\ding{173}}
        \psfig{file=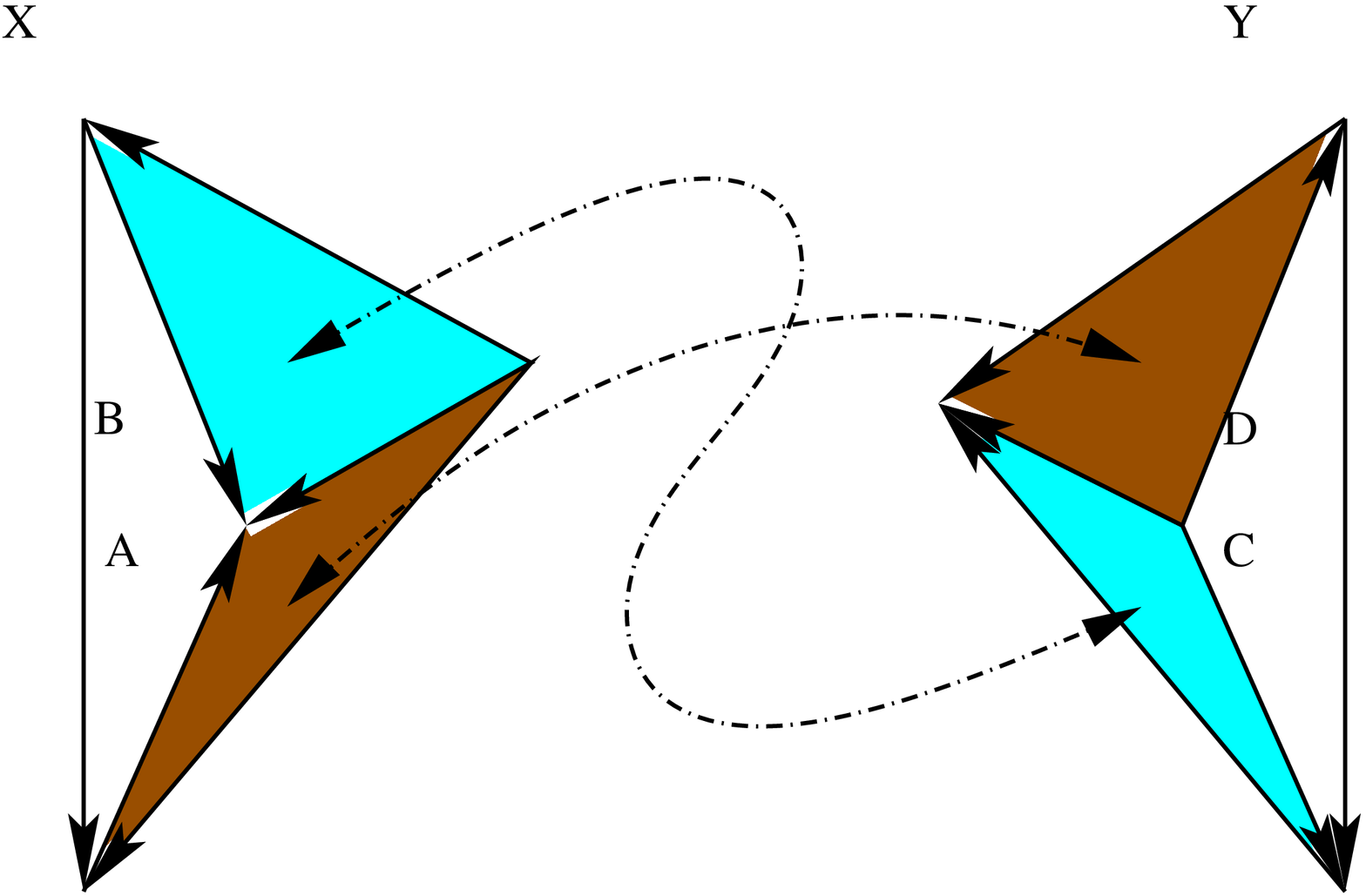,scale=0.32}
      \end{psfrags}
      }
    ~~~$\Longleftrightarrow$~~~
    \parbox{4cm}{
      \begin{psfrags}
        \psfrag{A}{$S^2_{0,4}$}
        \psfrag{B}{\textcolor{red}{\reflectbox{$S^1_{1,4}$}}}
        \psfrag{C}{\textcolor{red}{\reflectbox{$S^1_{0,4}$}}}
        \psfrag{D}{$S^2_{1,4}$}
        \psfig{file=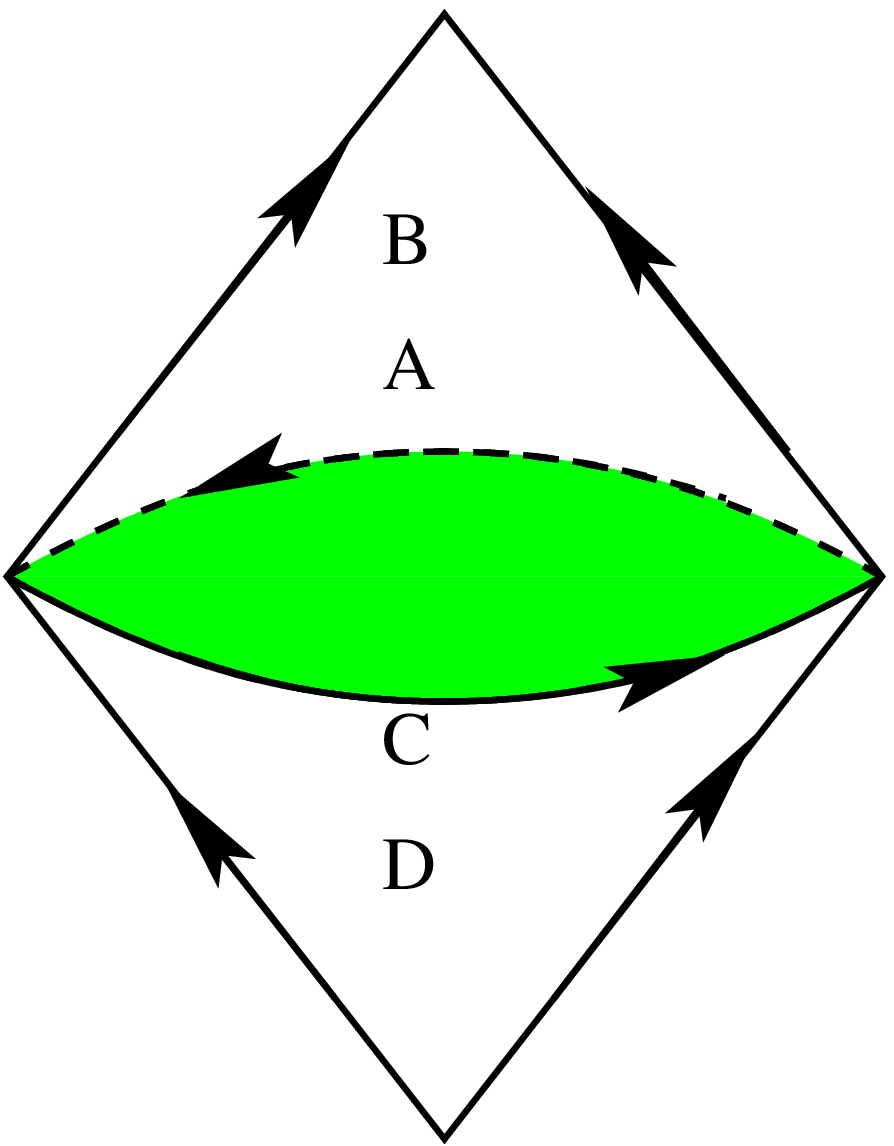,scale=0.36}
      \end{psfrags}
      }
    \caption{Gluing 2 tetrahedra in a surface $D_4$ gives a suspension.}
    \label{fig:41_glue_sp}
  \end{center}
\end{figure}

\begin{figure}[htbp]
  \begin{center}
    \begin{psfrags}
      \psfrag{A}{$S^4_{2,5}$}
      \psfrag{B}{\textcolor{red}{\reflectbox{$S^3_{3,5}$}}}
      \psfrag{C}{\textcolor{red}{\reflectbox{$S^3_{2,5}$}}}
      \psfrag{D}{$S^4_{3,5}$}
      \psfig{file=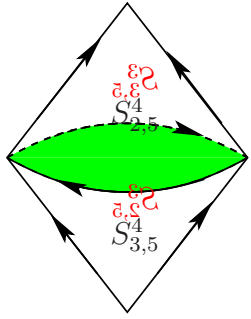,scale=0.36}
    \end{psfrags}
    \caption{Gluing 2 tetrahedra  in a surface $D_5$.}
    \label{fig:41_glue_sp_2}
  \end{center}
\end{figure}

In a surface $D_4$, we have 2 tetrahedra, which are glued to
each other
and result  in a suspension
as
shown in Fig.~\ref{fig:41_glue_sp}.
We also get a suspension from a surface $D_5$ as is shown in
Fig.~\ref{fig:41_glue_sp_2}.

\begin{figure}[htbp]
  \begin{center}
    \begin{psfrags}
      \psfrag{A}{\textcolor{red}{\reflectbox{$P_0$}}}
      \psfrag{B}{\rotatebox{30}{$P_1$}}
      \psfrag{C}{\textcolor{red}{\reflectbox{$P_2$}}}
      \psfrag{D}{\rotatebox{-20}{$P_3$}}
      \psfrag{E}{\textcolor{red}{\reflectbox{$P_0$}}}
      \psfrag{F}{\textcolor{red}{\reflectbox{$P_1$}}}
      \psfrag{G}{\rotatebox{-20}{$P_2$}}
      \psfrag{H}{\rotatebox{30}{$P_3$}}
      \psfig{file=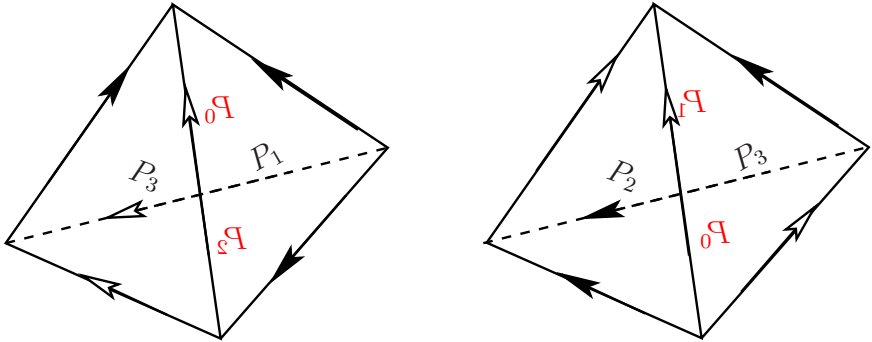,scale=0.5} 
    \end{psfrags}
    \caption{Complement of the figure-eight knot.}
    \label{fig:comple_41}
  \end{center}
\end{figure}

We next glue these polyhedra,
4 2-adjacent-tetrahedra (Fig.~\ref{fig:41_glue_1}
and Fig.~\ref{fig:41_glue_2}) and 2 suspensions
(Fig.~\ref{fig:41_glue_sp} and Fig.~\ref{fig:41_glue_sp_2}).
We first cut those polyhedra in the plane which was painted  gray in
figures, and separate them into ``upper'' and ``lower'' polyhedra
(tetrahedra or cones).
In this procedure, we should remember  which vertices
in the gray faces
were  glued to each other.
We then glue these polygons to each other
which have  same surfaces $S^a_{i,j}$,
and we finally obtain 2 tetrahedra
(Fig.~\ref{fig:comple_41}) which come from   upper and lower polyhedra.
Faces in these tetrahedra present  gray faces
in Figs.~\ref{fig:41_glue_1}--\ref{fig:41_glue_2} which separates
upper and lower polyhedra.
It is a well known result by Thurston~\cite{WPThurs80Lecture} that the
complement of
the figure-eight knot is decomposed into these 2 tetrahedra.
Note that,  in order to have upper and lower polyhedra, we have used the so-called
1-4 Pachner move which shows that
neighborhood of  vertex inside the tetrahedron  constitute sphere
as was pointed out in Ref.~\citen{YYokot00b}.

The partition function of the complement of the figure-eight knot is
then computed from Fig.~\eqref{fig:comple_41} as
\begin{align}
  {\tau}_1({4_1})
  & =
  \int \mathrm{d} \vec{p} \
  \langle p_1 = 0, p_2 \  | \  S \  | \  p_3, p_4 \rangle \,
  \langle p_4, p_3 \ | \  S^{-1} \  | \  p_2 , p_1=0\rangle
  \nonumber
  \\
  & \sim
  \int \mathrm{d} p \
  \exp \frac{1}{2 \, \mathrm{i} \, \gamma} \,
  \Bigl(
    \Li (\mathrm{e}^{-p}) - \Li (\mathrm{e}^p)
  \Bigr) .
\end{align}
Here we have introduced a restriction $p_1=0$ which comes from a
computation of the knot invariant for a ($1,1$)-tangle.
The integral in the partition function can be evaluated at the saddle
point,
\begin{equation*}
  (1- \mathrm{e}^p) \, ( 1 - \mathrm{e}^{-p})=1 ,
\end{equation*}
which, with a root of $\omega^2 - \omega + 1=0$, gives
\begin{equation}
  \Tilde{\tau}_1(4_1)
  =
  \lim_{\gamma \to 0}
  \left(
    2 \, \mathrm{i} \, \gamma \log {\tau}_1(4_1)
  \right)
  =
  2.02988 \, \mathrm{i} .
\end{equation}
The imaginary part is nothing but the hyperbolic volume of the
complement of the figure-eight knot.

\section{Concluding Remarks}

We have studied the knot invariant
by use of the infinite dimensional representation of the quantum
dilogarithm function.
This invariant can be seen
by construction  as the non-compact analogue
of the colored Jones polynomial.
We  have found  that,
by assigning the oriented tetrahedra to the $S$-operator
which solves the five-term relation,
the braid operators can be  depicted  as the octahedron as was
shown  in Ref.~\citen{DThurs99a}.
With this realization,
we have obtained a general scheme to triangulate the knot complements.
This method can be applicable   for arbitrary knots and links, whereas
methods in Refs.~\citen{MoTaka85a,CPetro97a} seem to  work 
only  for the
alternating knots.
We have further
revealed that the hyperbolic structure appears   in the
classical limit of our invariant, and that
the hyperbolicity consistency
conditions in gluing ideal tetrahedra  coincide exactly
with the saddle point  equations of integrals of knot invariant.
Based on the result  that
an imaginary part of the
$S$-operator~\eqref{define_S_operator}
reduces to the Bloch--Wigner function at the critical
point~\eqref{property_V},
and that we can identify the $S$-operator as the oriented ideal
tetrahedron whose dihedral angles are  fixed,
we  can  conclude that the imaginary part of the invariant
$\Tilde{\tau}_1(K)$  will give the hyperbolic volume
\begin{align*}
%
  \Im \, \Tilde{\tau}_1(K)
  &=
  \sum_{\text{ideal tetrahedra}:  i}
  D(z_i) ,
\end{align*}
though we are not sure which solutions of a set of the hyperbolicity
conditions are dominant in the classical limit.

{}From the physical view points, the Jones polynomial
is closely related with the
topological gauge field theory in 3-dimension~\cite{EWitt89a}.
Therein the Chern--Simons path integral becomes the invariant of the
3-manifold, and in
Ref.~\citen{DijkWitt90b}
Dijkgraaf and Witten gave a combinatorial definition
for the Chern--Simons invariants $\mathrm{CS}(M)$
of the manifold $M$
by use of 3-cocycles of the group
cohomology~\cite{DijkWitt90b}.
What is interesting here is that,
with  the hyperbolic volume $\mathrm{Vol}(M)$,
the function
$\mathrm{Vol}(M) + \mathrm{i} \, \mathrm{CS}(M)$
is analytic and depends on
element $\beta(M)$ of  the  (orientation sensitive) scissors congruence
group~\cite{NeumZagi85a,TYoshi85a}.
With  a suitable setting of   branches of the log
function,
the element of the  scissors congruence group  for non-compact
manifold $M$ is known
to be given by~\cite{NeumYang95b}
\begin{equation}
  \beta(M)
  =
  \sum_{
    \text{ideal triangulation $z_i$}
    }
  L(1 - z_i) ,
\end{equation}
where
$L(z)$ is the Rogers dilogarithm function~\eqref{Rogers}.
As we have shown that
the invariant $\Tilde{\tau}_1(K)$ is given by eq.~\eqref{result_tau}
based on that
the $S$-operator in the classical  limit gives the Rogers dilogarithm
function at the critical  point~\eqref{property_V},
it is   a natural consequence that our non-compact colored Jones
invariant~\eqref{limit_invariant} may
give the Chern--Simons term,
\begin{equation}
  \Tilde{\tau}_1(K)
  =
  \lim_{\gamma \to 0}
  \Bigl(
  2 \, \mathrm{i} \,  \gamma
  \log \tau_1(K)
  \Bigr) 
  =
  \mathrm{i} \,
  \Bigl(
  \mathrm{Vol}(K)
  + 
  \mathrm{i} \, \mathrm{CS}(K)
  \Bigr) .
\end{equation}



\section*{Acknowledgement}
The author would like to thank Hitoshi Murakami for  kind explanation
of his results and for
stimulating discussions.
He is also grateful to R.~Kashaev and Y.~Yokota for useful communications.
Thanks are to Referee for useful comments.

\newpage
\bibliographystyle{physics}

\end{document}